\documentclass[journal]{IEEEtran}

\usepackage{color}

\usepackage{graphicx}
\usepackage{algorithm}
\usepackage[noend]{algpseudocode}
\usepackage{amsmath}
\usepackage{amssymb}
\usepackage[export]{adjustbox}
\usepackage{subfigure}
\usepackage{float}
\usepackage[flushleft]{threeparttable}
\usepackage{multirow}
\usepackage[switch]{lineno}
\usepackage{array}

\usepackage[english]{babel}
\usepackage[dvipsnames,svgnames,x11names]{xcolor}
\usepackage[markup=underlined]{changes}

\definechangesauthor[color=BrickRed]{I}
\definechangesauthor[color=NavyBlue]{JH}

\usepackage{todonotes}
\makeatletter
\setremarkmarkup{\todo[color=Changes@Color#1!20,size=\scriptsize]{#1: #2}}
\makeatother

\ifCLASSINFOpdf
 
\else

\fi

\begin{document}

\title{Hierarchical Temporal Memory using Memristor Networks: A Survey}


\author{Olga Krestinskaya,~\IEEEmembership{ Graduate Student Member, IEEE,}
        Irina Dolzhikova,~~\IEEEmembership{Graduate Student Member,~IEEE},  and Alex Pappachen James,~\IEEEmembership{Senior Member,~IEEE}  
\thanks{A.P. James is a faculty and director of the Bioinspired Microelectronics Systems Lab with the School of Engineering, Nazarbayev University; Corresponding author, e-mail: (see http://www.biomicrosystems.info/alex).}
\thanks{O. Krestinskaya and I. Dolzhikova are graduate students and research assistants with the Bioinspired Microelectronics Systems Group, Nazarbayev University}
}

\maketitle

\begin{abstract}
This paper presents a survey of the currently available hardware designs for  implementation of the human cortex inspired algorithm, Hierarchical Temporal Memory (HTM). In this review, we focus on the state of the art advances of memristive HTM implementation and related HTM applications. With the advent of edge computing, HTM can be a potential algorithm to implement on-chip near sensor data processing. 
The comparison of analog memristive circuit implementations with the digital and mixed-signal solutions are provided.
The advantages of memristive HTM  over digital implementations against performance metrics such as processing speed, reduced on-chip area and power dissipation are discussed. The limitations and open problems concerning the memristive HTM, such as the design scalability, { sneak currents, leakage,} parasitic effects, lack of the analog learning circuits implementations and unreliability of the memristive devices {integrated with CMOS circuits} are also discussed.
\end{abstract}

\begin{IEEEkeywords}
Hierarchical Temporal Memory, Spatial Pooler, Temporal Memory, Memristor, Spin-neuron, Crossbar. 
\end{IEEEkeywords}

\IEEEpeerreviewmaketitle

\section{Introduction}

\IEEEPARstart{H}{ierarchical} temporal memory (HTM) is a neuromorphic machine learning algorithm that emulated the performance of the human brain neocortex. The main characteristics of HTM are sparsity, hierarchy and modularity. HTM consists of two main parts: the HTM Spatial Pooler (SP) and the HTM Temporal Memory (TM). The HTM SP converts the inputs to sparse binary patterns and produces Sparse Distributed Representation (SDR) of the input data. This process also refers to the encoding of the information, which is performed throughout the HTM SP. The HTM SP is useful for visual data processing, involving recognition {\cite{doremalen2008spoken, maltoni2011pattern, mattsson2011fruit}}, categorization {\cite{csapo2007object}} and classification {\cite{perea2009application}} problems. The HTM TM is the HTM part responsible for the learning process.  


Algorithmic implementations of the HTM on software revealed great potential of this machine learning method for large variety of applications. The increase in the edge computing devices in the Internet of things era drives the need for hardware acceleration for near sensor processing and computing. The computational considerations of the processing speed and possibility for the real-time realization pushed for transition of the originally purely software based algorithm to the hardware implementation. As a result, hardware design of HTM became a highly attractive topic for {the} last few years that already produced some promising results. 

In this work, we provide a comprehensive review of the analog memristive HTM implementations, which are derived and modified from the original HTM algorithm. We present the state-of-the-art studies involving the implementation of memristive HTM. We draw a comparison between the HTM digital solutions and analog memristive HTM implementations and analyze the advantages, drawbacks and open problems. We describe and compare the existing HTM solutions in relation to various applications, comparing the performance accuracy, on-chip and power dissipation. We discuss the scalability issues, the possibility and drawbacks of real memristive HTM implementation. We provide a summary of the open problems concerning memristive HTM implementation that should be addressed.

This paper is organized into 6 sections. Section \ref{s2} describes the HTM theory and introduces the mathematical framework for the HTM SP and TM algorithms. Section \ref{s3} provides the comprehensive review of the hardware implementations of HTM with the focus on the mixed-signal and purely analog implementations based on the memristive circuits. Section \ref{s4} represents the survey on the HTM applications including HTM performance, power and area calculations and overview of the system level HTM implementation process. 
Section \ref{s5} provides the explicit discussion of the advantages of the memristive HTM, limitations of the hardware design and open problems. Section \ref{s6} concludes the paper.

 

\section{HTM Theory}
\label{s2}
\subsection{HTM Overview}

HTM is a machine learning algorithm that is inspired by the biological functionality of the neocortex \cite{hawkinsintelligence}. This algorithm attempts to reproduce the following characteristics of the neocortex: processing and representation of the information in the sparsely distributed manner, encoding of the sensory data in real time  \cite{bami2016},  consideration of the temporal changes of the input data and ability to make predictions based on the history of previous events \cite{george2005hierarchical}.

The implementation of HTM requires consideration of two modules: Spatial Pooler (SP) and Temporal Memory (TM). The studies on functionality of HTM revealed that SP alone is able to learn and classify the data sets that are related to the numbers, pixels and letters \cite{zyarah2015design}. This was validated on the practical applications related to the biometric recognition, object categorization \cite{csapo2007object}, face recognition \cite{ibrayevdesign,
fedorova2016htm, tcad}, speech recognition \cite{fedorova2016htm}, gender classification \cite{james2017design}, handwritten digits recognition \cite{fan2016hierarchical, zyarah2015design, streat2016non} and natural language processing \cite{poster}. 
The purpose of SP in HTM is to train the system to recognize certain patterns  of the input data such that for different inputs  that possess similar characteristics particular attributes in the output are activated \cite{hawkinsintelligence}. 
In other words, SP generates the sparse distributed representation (SDR) from the given input data. While SP considers the common patterns in the spatial domain, TM looks for the temporal patterns. It examines the temporal changes of the input data and makes predictions based on previous experience of having certain patterns being followed by particular types of other patterns {\cite{25}}.

\begin{figure}[!h]
\centering
\includegraphics[width=3.3in]{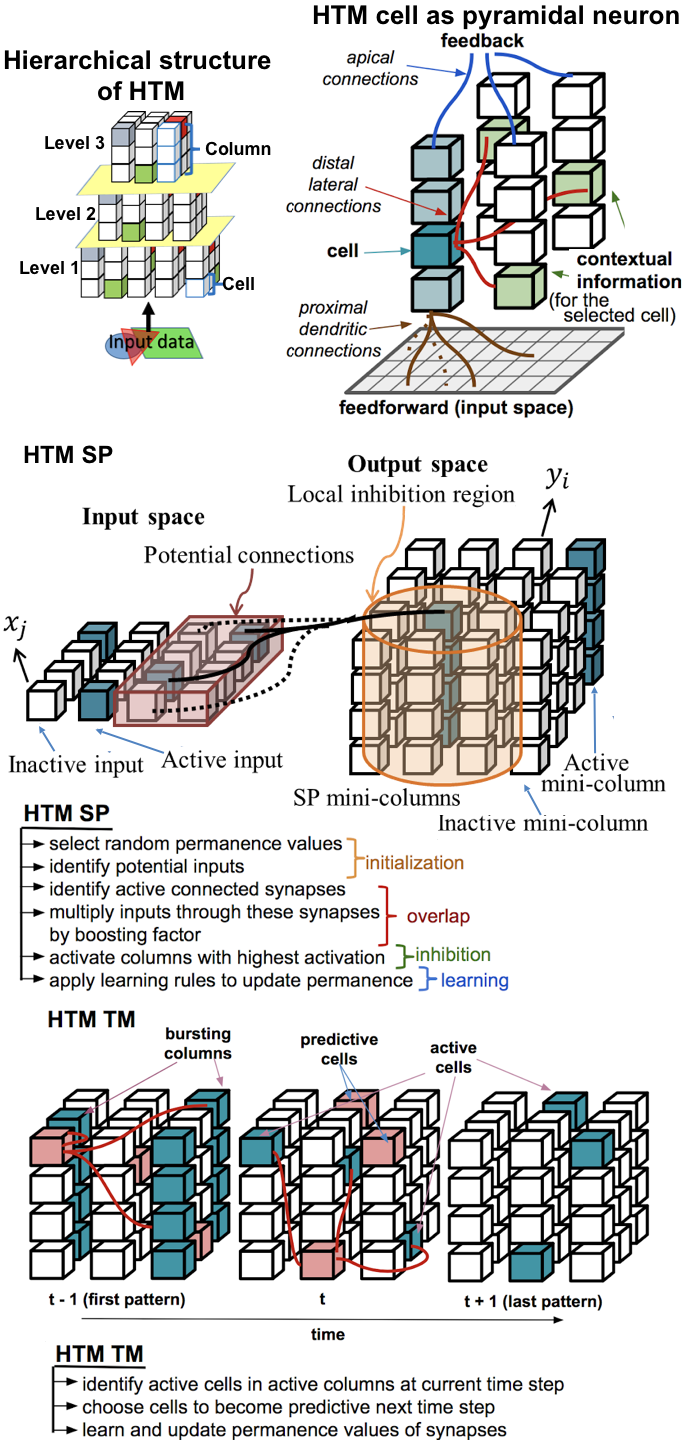}
\caption{HTM overview containing 3-level hierarchical structure of HTM (top left), HTM representation related to biological concepts of pyramidal neuron (top right), HTM SP and HTM TM main structure and brief explanation of their main steps. }
\label{omg}
\end{figure}

{ The 3-level hierarchical structure of HTM is shown in Fig. \ref{omg} (top left). The fundamental elements within the hierarchy are cells that are grouped together to form a column. The cells in the hierarchy of HTM are the basic processing units that are used to emulate the functionality of the pyramidal neurons, where the output depends not only on the feedforward input, but also feedback and contextual information learned from previous inputs (Fig.\ref{omg}, top right). The group of columns defines a region within the HTM hierarchy. Although 3-level hierarchy in Fig. \ref{omg} (top left) illustrates a single region within each of the levels, depending on the application there might be several regions. The level that is higher in the hierarchy uses the patterns from the lower levels, which usually represents more details and consequently requires more regions for the processing of these details.}  
 The 3-level hierarchy in Fig. \ref{omg} (top left) also illustrates an idea of how the analog data is transformed to the SDR. The information from the input data being the geometrical figures of different colors is transformed to the activation of certain cells in the regions of HTM.

%
The realization of SP requires implementation of four phases, which are 
initialization, calculation of overlap value, inhibition and learning process \cite{hawkinsintelligence, spnew}. During initialization phase certain number of columns within the HTM regions are selected. The selected columns are considered to receive the input data. The connection between the input and the column is established through the dendrite segment with several potential synapses. The initialization of the weight values of the potential synapses (permanence values) is performed randomly. Those potential synapses that obtained the permanence value greater than threshold are said to be connected. The activity of the synapse is related to the input whether it is active (high input) or not. 
 {The aim of the overlap phase is to compute} 
 the number of $active$ $connected$ synapses 
 \cite{bami2016}. The frequency of the column being active with respect to its neighbors is considered by the boosting factor that is used to multiply with the active synapses. 
 During the inhibition phase the columns that are inside of the same inhibition region having $k$-th highest activations become active, while others are inhibited. During the learning phase the rule for the Hebbian learning is used to update the weights of the synapses. The phases of overlap, inhibition and learning are then repeated. 
 
 The outputs that are collected from SP are used as inputs to TM. The active columns of SP produces the so-called feed-forward inputs. The first step for the TM implementation is to activate the cells that are in the predictive state within the columns that won during the inhibition stage (active columns). In case if there are no cells that are in predictive state, all of the cells within the column are activated. For each of the dendrite segments the number of the synapses that are connected to the active cells is computed. If the number of the synapses is higher than the given threshold the segment is considered being active. The cells that are related to the activated dendrite segment are settled to the predictive state. The prediction in TM is based on consideration of the cells that in the predictive state. The weights of the potential synapses that are within the activated dendrite segment are updated based on the activity of the cell. The active cell leads to the increase in the permanence value. This is considered to be a temporal change that is applied only if the feed-forward activation of the cell is correct. In case if the cell makes incorrect prediction on the feed-forward activation of the other cell, the temporal changes are removed and 
 permanence value of the synapse being active is reduced. 
 
 
 

\subsection{Mathematical Framework}

\subsubsection{Spatial Pooler}


{An arrangement of the input space and the output space of the HTM SP and its main steps are shown in Fig. \ref{omg}}, where $x_j$ refers to the $j$-th input neuron in the input space. The outputs are topologically arranged into mini-columns, where $y_i$ is the $i${-}th output SP mini-column in the output space. The mini-column $y_i$ is connected to the part of the input space, which is called potential connections. The synapse refering to the $i$-th SP mini-column is located in a hypercube of the input space with the center in  $x_i^c$ and edge length $\gamma$. If the synaptic permanence of the synapse is greater than the threshold value, the synapse is  connected. Eq.\ref{olgaeq1} represents the potential input connections $PI(i)$ between to the input space and $i$-th mini-column \cite{25,tcad}.

\begin{equation}
PI(i)= \left \{ j|\imath(x_j;x_i^c,\gamma)  \;\textup{and} \; (z_{ij}<\rho) \right \}
\label{olgaeq1}
\end{equation}

where, $\imath(x_j;x_i^c,\gamma)=1, \forall x_j\in (x_i^c,\gamma)$, and $z_{ij}\sim U(0,1)$, $z$ is selected randomly from the uniform distribution $U$ having range $[0,1]$. The parameter $\rho$ denotes the fraction of inputs that are potential connections within the hypercube of the input space. {The uniform distribution is used for the synaptic performances to ensure that all of the synapses have equal probability of being connected \cite{mnatzaganian2017mathematical}.}

The set of connected synapses are represented as a binary matrix $\mathbf{B}$ in Eq. \ref{eqolga2}. The synapses are considered to be connected, if their value is above the threshold value $\theta_c$. 

\begin{equation}
\mathbf{B}_{ij}= \left\{\begin{matrix}
 1& \textup{if} \quad \mathbf{S_{ij}}\geq \theta_c \\ 
 0& \textup{otherwise}
\end{matrix}\right.
\label{eqolga2}
\end{equation}

The threshold parameter  $\theta_c$ refers to the percentage of connected synapses. For instance, $\theta_c=0.4$ indicates that 40\% of the potential synapse are connected. The synaptic permanence from the $j$-th input to the $i$-th SP mini-column is represented by the matrix $\mathbf{S_{ij}} \in [0,1]$ shown in Eq. \ref{olgaeq3}.  

\begin{equation}
\mathbf{S}_{ij}= \left\{\begin{matrix}
 U(0,1)& \textup{if} \quad j\in PI(i)\\ 
 0& \textup{otherwise}
\end{matrix}\right.
\label{olgaeq3}
\end{equation}

Next, the local inhibition part of the HTM SP is executed through the local inhibition mechanism that implies the SP mini-columns in a particular neighborhood inhibit each other. This neighborhood $\mathbf{N}_i$ of the $i$-th SP mini-column is defined by Eq. \ref{olgaeq4}, where, $\left \| y_i-y_j \right \|$ determines the Euclidean distance between the mini-columns $i$ and $j$ and the parameter $\phi$ is used to control the inhibition radius.

\begin{equation}
\mathbf{N}_i= \left\{ j|\left \| y_i-y_j \right \|<\phi, i\ne j \right\}
\label{olgaeq4}
\end{equation}

 The parameter $\phi$ is useful when the inputs are also topologically arranged to the mini-columns to ensure that all the inputs are affected by the inhibition process. If the receptive field size increases, the parameter $\phi$ also increases. To determine $\phi$, the average number of connected input spans of all the SP mini-columns are multiplied by the number of mini-columns per inputs. If the dimensions of SP inputs and mini-columns are same, $\phi$ is set equal to $\gamma$.
 
The measure that determines the activation of SP mini-columns for a given input pattern $\mathbf{Z}$ is called input overlap. The input overlap shown in Eq. \ref{olgaeq5} is calculated as a feed-forward input to each mini-column.

\begin{equation}
o_i= \beta_i \sum_j \mathbf{B}_{ij}\mathbf{Z}_j
\label{olgaeq5}
\end{equation}

The parameter $\beta_i $ represents the boosting factor that controls the excitability of each SP mini-column. The appropriate value of the boosting factor is either selected initially or modified during training and learning phase of the HTM SP. 

The activation of the SP mini-column depends on two main conditions: the input overlap is greater than a stimulus threshold $\theta_s$ and is among the top $s$ percentile ($prctile$) of its neighborhood. Eq.\ref{olgaeq6} refers to the  active column selection, where $\alpha_i$ is the activity of the SP mini-columns, and $\mathbf{NO}(i)=\left\{o_j|j\in \mathbf{N}(i)  \right\}$ with $s$ being the target activation density.  This activation rule refers to the implementation of the $k$-winners-take-all
computation within a local neighborhood.

\begin{equation}
\alpha_i= 1, \textup{if} \quad (o_i \geq \textup{prctile}(\mathbf{NO}(i),1-s)) \;\textup{and}\; (o_i\geq \theta_s)
\label{olgaeq6}
\end{equation}

 In the original algorithm, the parameter $k$, can be adjusted to regulate the desired number of winning columns \cite{25}. However, to simplify the algorithm for the {circuit level implementation},  the value of the desired activity level is limited to $1$ because the the inhibition phase is implemented by the Winner-Takes-All (WTA) circuits \cite{ibrayevdesign, fedorova2016htm}.

In the learning phase of the HTM SP, feed-forward connections are learned using Hebbian rule and the boosting factor is updated considering the last $T$ inputs in time. The reinforcement of input connections is performed by the increasing of the permanence value $\rho$ by $\rho ^+$. Whereas, the permanence of inactive connections is decrease by $\rho ^-$. It should be noticed that the permanence values are kept withing the boundaries of from $0$ to $1$.

The update process of the boosting factor depends on the time-average activity and the recent activity of the SP mini-columns \cite{25}. The parameter $\bar{\alpha_i} (t)$ refers to the time-average activation level in time $t$ considering $T$ previous inputs and current activity $\alpha_i (t)$ of the $i$-th mini-column shown in Eq.\ref{olgaeq66}.

\begin{equation}
\bar{\alpha_i} (t)= \frac{(T-1)\times \bar{\alpha_i}(t-1)+\alpha_i (t)}{T}
\label{olgaeq66}
\end{equation}

The calculation of the recent activity $<\bar{\alpha_i} (t)>$ of the mini-columns in a particular neighborhood in time $t$ is shown in Eq.\ref{olgaeq7}.

\begin{equation}
<\bar{\alpha_i} (t)>= \frac{1}{|\mathbf{N}(i)|} \displaystyle\sum_{j\in \mathbf{N}(i)}^{} \bar{\alpha_i} (t)
\label{olgaeq7}
\end{equation}

The update process of the boosting factor is shown in Eq.\ref{olgaeq8}, where the parameter $\eta$ controls the adaptation effect in the HTM SP. 

\begin{equation}
\beta_i (t)= e^{-\eta(\bar{\alpha_i} (t)-<\bar{\alpha_i} (t)>)}
\label{olgaeq8}
\end{equation}

\subsubsection{Temporal Memory}

{ HTM TM implementation and sequence learning process is illustrated in Fig. \ref{omg}.} The predictive state of the cells in the HTM TM is shown in Eq.\ref{olgaeq9}, where $A^t$ is the activation matrix of size $N\times S$ refering to $N$ columns and $S$ neurons per column with elements $a_{ij}^t$ of $A^t$ denoting the activation state of the $i$-th cell and $j$-th column at the time point $t$. The permanence of the $d$-th segment of $i$th cell in the $j$-th column is denoted as $D_{ij}^d$. The parameter $\widetilde{D}_{ij}^d$ corresponds to connected synapses, and $\theta$ refers to the segment activation threshold value \cite{tmnew}.

\begin{equation}
\pi_{ij}^t=\begin{cases}
 1 & \text{ if } \exists_d \left \| \widetilde{D}_{ij}^d \cdot A^t \right \|_1 >\theta \\ 
 0 &  \text{ Otherwise }
 \label{olgaeq9}
\end{cases}
\end{equation}

 The calculation of the activation state is shown in Eq. \ref{olgaeq10}, where, $W^t$ refers to the top $s1$ percentage of column that has largest number of synaptic inputs. Typically $s1$ is set to 1\% to 2\% \cite{tmnew}. The cell in the winning column is activated, {if} it was in the {predictive state} in the previous time step.
 
\begin{equation}
a_{ij}^t=\begin{cases}
 1 & \text{ if } j \in W^t \text{ and } \pi_{ij}^{t-1}=1\\ 
  1 & \text{ if } j \in W^t \text{ and } \sum_i\pi_{ij}^{t-1}=0\\ 
 0 &  \text{ Otherwise }
\end{cases}
\label{olgaeq10}
\end{equation}

The dendrite segment activations and the learning of lateral connections is preformed using a Hebbian-like rule. Eq. \ref{olgaeq11} represents the reinforcement of the depolarized cell in a segment that subsequently become active, where $\dot{D}_{ij}^d$ is a binary matrix representing only positive elements of $D_{ij}^d$, and the small values $\rho ^-$  and $\rho ^+$ are for negative reinforcement of inactive synapse and positive reinforcement of active synapse in dendrite  segments, respectively.
\begin{equation}
\Delta D_{ij}^d=\rho ^+ \dot{D}_{ij}^d \cdot A^{t-1}-\rho ^- \dot{D}_{ij}^d \cdot (1-A^{t-1})
\label{olgaeq11}
\end{equation}

 Eq. \ref{olgaeq12} refers to the long term that is introduced by including a small decay to the active segments that did not become active,  where $a_{ij}^t=0$ and $\left \| \widetilde{D}_{ij}^d \cdot A^{t-1} \right \|_1>\theta$, with $ \widetilde{\rho} ^- <<\rho ^-$ \cite{tmnew, tcad}.
\begin{equation}
\Delta D_{ij}^d= \widetilde{\rho} ^- \dot{D}_{ij}^d 
\label{olgaeq12}
\end{equation}

The research study \cite{tmnew} states that the Hebbian-like learning rule can be replace{d} by the gradient descent method on the cost function, such as prediction error, which could lead to the better accuracy results. {However, the number of studies of the HTM implementation with backpropagation with gradient descent for weight update is limited, and the performance accuracy comparison of these two methods has not been done yet. }

\begin{algorithm}
\caption{HTM algorithm}\label{alg}
\begin{algorithmic}[1]


\State {Define the size of input neighborhood with potential connections, $x_i^c$, $\gamma$, $\rho$, $\eta$, $\theta_c$, size of the local inhibition region, $\theta_s$}
\Comment{\textbf{HTM SP}}
\State{Determine $\phi$ by multiplying the average number of connected input spans of all the SP mini-columns by the number of mini-columns per inputs.}

 \If{size (inputs)=size(mini-columns)} 
 \State $\phi=\gamma$
 \EndIf

 \State $z_{ij}\sim U(0,1)$

 \If{$\forall x_j\in (x_i^c,\gamma)$} 
 \State $\imath(x_j;x_i^c,\gamma)=1$
 \EndIf

 \For{$\imath(x_j;x_i^c,\gamma)  \;\textup{and} \; (z_{ij}<\rho)$} 
 \State $PI(i)= j$
 \EndFor

 \If{$j\in PI(i)$} 
 \State $\mathbf{S}_{ij}= U(0,1)$
 \Else
 \State $\mathbf{S}_{ij}=0$
 \EndIf

 \If{$\mathbf{S_{ij}}\geq \theta_c$} 
 \State $\mathbf{B}_{ij}=1$
 \Else
 \State $\mathbf{B}_{ij}=0$
 \EndIf
 
 \For{$| y_i-y_j |<\phi, i\ne j $} 
 \State $\mathbf{N}_i=  j$
 \EndFor

\State{$o_i= \beta_i \sum_j \mathbf{B}_{ij}\mathbf{Z}_j$}

 \For{$j\in \mathbf{N}(i) $} 
 \State $\mathbf{NO}(i)=o_j$
 \EndFor

 \If{$(o_i \geq \textup{prctile}(\mathbf{NO}(i),1-s)) \;\textup{and}\; (o_i\geq \theta_s)$} 
 \State $\alpha_i= 1$
 \Else
 \State $\alpha_i= 0$
 \EndIf

\If {input connections are active}
\State$\rho=\rho+\rho ^+$
\ElsIf {input connections are inactive}
\State$\rho=\rho+\rho ^-$
\EndIf

\For {time period $t$}
\State{$\bar{\alpha_i} (t)= \frac{(T-1)\times \bar{\alpha_i}(t-1)+\alpha_i (t)}{T}$}

\EndFor
 
\State{$<\bar{\alpha_i} (t)>= \frac{1}{|\mathbf{N}(i)|} \displaystyle\sum_{j\in \mathbf{N}(i)}^{} \bar{\alpha_i} (t)$}

\State{$\beta_i (t)= e^{-\eta(\bar{\alpha_i} (t)-<\bar{\alpha_i} (t)>)}$}

\State {Define $N$, $S$, $\theta$}
\Comment{\textbf{HTM TM}}
 \If{$\exists_d \left \| \widetilde{D}_{ij}^d \cdot A^t \right \|_1 >\theta$} 
 \State $\pi_{ij}^t=1$
 \Else
 \State $\pi_{ij}^t=0$
 \EndIf
 
\If {$j \in W^t \text{ and } \pi_{ij}^{t-1}=1$}
\State$a_{ij}^t=1$
\ElsIf {$j \in W^t \text{ and } \sum_i\pi_{ij}^{t-1}=0$}
\State$a_{ij}^t=1$
\Else
\State{$a_{ij}^t=0$}
\EndIf 
  
\State{$\Delta D_{ij}^d=\rho ^+ \dot{D}_{ij}^d \cdot A^{t-1}-\rho ^- \dot{D}_{ij}^d \cdot (1-A^{t-1})$} 
 
 \If {$a_{ij}^t=0$ and $\left \| \widetilde{D}_{ij}^d \cdot A^{t-1} \right \|_1>\theta$, with $ \widetilde{\rho} ^- <<\rho ^-$}
\State$\Delta D_{ij}^d= \widetilde{\rho} ^- \dot{D}_{ij}^d$
\EndIf 
\end{algorithmic}
\end{algorithm}

The Algorithm \ref{alg} summari{z}es the original HTM structure and illustrates the main processing steps. Lines 1- 34 represents the HTM SP implementation, while lines 35-48 refer to the HTM TM processing. Lines 1-19 refer to the initialization stage and local inhibition calculations. Line 20 refers to the overlap calculations. The lines 21-26 represent the inhibition stage of HTM SP. Lines 27-30 includes the learning stage of the HTM SP, where reinforcement of active connections and suppression of inactive connections is performed. Lines 31-34 correspond to the update of boost factor. The HTM TM processing starts at line 35, where the number of columns and number of neurons per column is initialized the the threshold $\theta$ is defined. Lines 36-39 refer to the determination of predictive state of the cells. The calculation of the activation function state is performed in lines 40-45. Finally, the reinforcement of depolarization cell corresponds to line 46, and the long-term depression is shown in lines 47-48.

\section{Hardware Implementation}

\label{s3}

\subsection{Mixed-signal HTM Implementation} 
The mixed signal implementation of HTM is represented in \cite{fan2016hierarchical}. The HTM architecture is based on the memristive crossbar arrays combined with spin-neuron devices, spin-neuron based successive approximation register analog-to-digital converters (SAR ADCs) and Winner-Takes-All (WTA) circuits. The analog part of HTM comprised of dot-product multiplication is performed using the memristive crossbars during the inference stage to perform the pattern classification problem. The digital part of HTM involves the offline training stage, which is performed using external software. In the inference stage, the digital inputs are fetched into the crossbar and the analog crossbar outputs are converted back into digital {signals}.

\begin{figure}
\centering
\includegraphics[width=90mm]{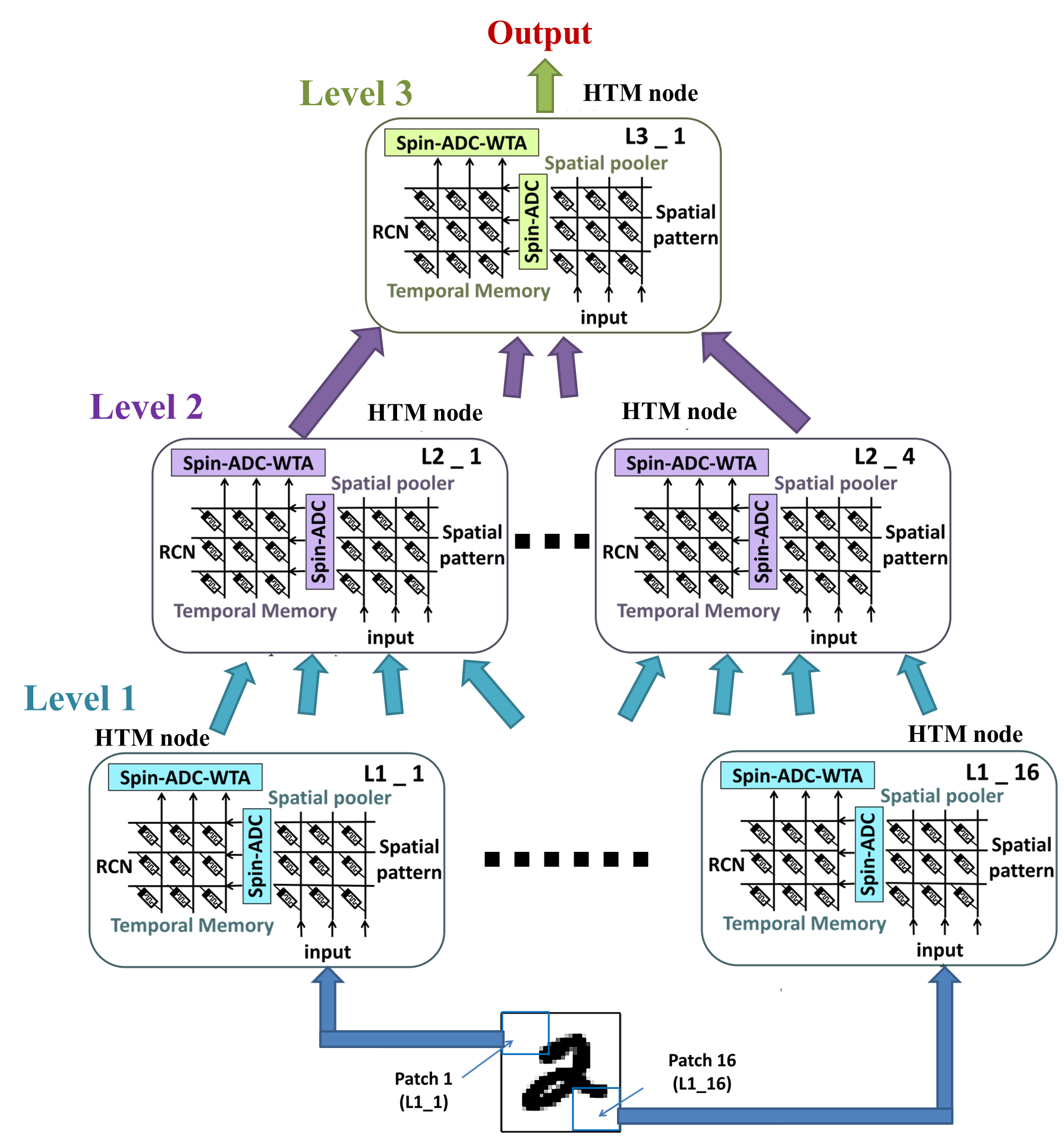}
\caption{Overall architecture of the mixed-signal design of HTM proposed in \cite{fan2016hierarchical}. The HTM architecture is hierarchical and modular. Each level of the hierarchy involves both analog and digital processing. This modular architecture is tested on handwritten digits recognition, where the parts of the patterns are fetched into the separate HTM node.}
\label{olgafig2}
\end{figure}

The overall system architecture is shown in Fig. \ref{olgafig2}. The system is tested for the handwritten digits recognition. The implemented HTM architecture is  hierarchical and modular. Each level of hierarchy contains certain number of HTM nodes, where each node includes the HTM SP, HTM TM and WTA circuits. Initially, a digital input pattern is divided into 16 patches and each patch is fetched into separate resistive crossbar networks (RCNs) at level 1. 

{The authors mentioned two possible ways to overcome the sneak path problem. One of the approaches is related to the application of the access transistors and diodes that was proposed in \cite{manem2011read}. 
The second method that was actually selected by the authors for the design of HTM circuit
does not require access transistors. This method suggests to access one memristive device at a time by selecting specific  column \cite{jung2012two}. However, this method could be used only if the programming speed is not of a primary importance.}

\begin{figure}
\centering
\includegraphics[width=90mm]{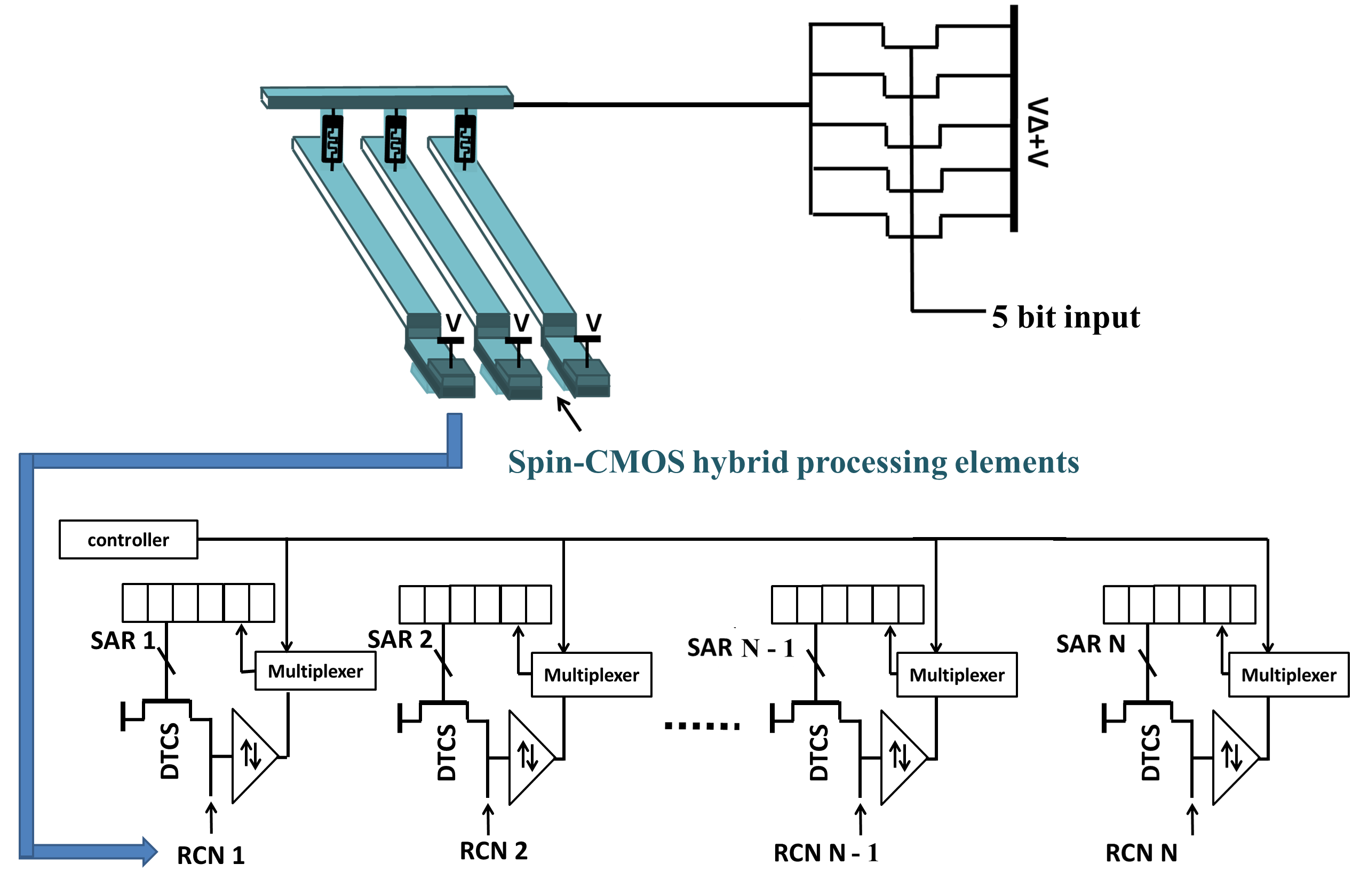}
\caption{RCN and SAR ADC architectures. RCN involves DTCS DAC and memristive crossbar. SAR ADC works as a current comparator and involves spin-neuron devices \cite{fan2016hierarchical}. }
\label{olgafig3}
\end{figure}

The RCN shown in Fig. \ref{olgafig3} consists of a Deep Triod Current Source Digital-to-Analog Converter (DTCS DAC) and memristive crossbar.
RCNs convert the digital inputs into the analog form. Before the crossbar multiplication, the inputs are converted into the analog form by the DTCS DAC in the switching circuit. Next, the DTCS DAC analog output signals are processed by the HTM SP. The crossbar analog outputs are represented as the output current from each crossbar column. The output current from HTM SP crossbars are detected and converted into digital form using spin-neuron based SAR ADC. The performance of spin-neuron devices is akin to the current-mode comparator because these devices switch when the current greater than the certain threshold is flowing through it. In the SAR ADC, the digital values stored in the approximation register are converted into analog currents using DTCS DAC. These currents are compared to the currents from the memristive crossbar. Then, a special latch is used to detect the output stage. Next, the SAR ADC digital outputs are fetched into the HTM TM RCNs having the same operating principles as HTM SP RCN. Finally, the digital output of HTM TM RCNs are fetched to the WTA circuit, which identified the winning index. If the HTM node is the highest in the hierarchy, the identified winning index refers to the class of the pattern. Otherwise, it represents the index of a  particular temporal group.

\subsection{Analog Implementation}
\subsubsection{Memristor-CMOS Hybrid Implementation of SP}

One of the first works that presented the analog implementation of Spatial Pooler is based on memristor-CMOS hyrbrid circuit \cite{ibrayevdesign}. 
Similar to any neural circuit, SP design requires implementation of the synapse that allows for communication between the neurons. Based on this requirement, the work in \cite{ibrayevdesign} demonstrates the design for the single synapse circuit. This design is then used to construct the column of SP that consists of $S$ number of such synapses.

\begin{figure}[!ht]
\centering
\includegraphics[width=1.8in]{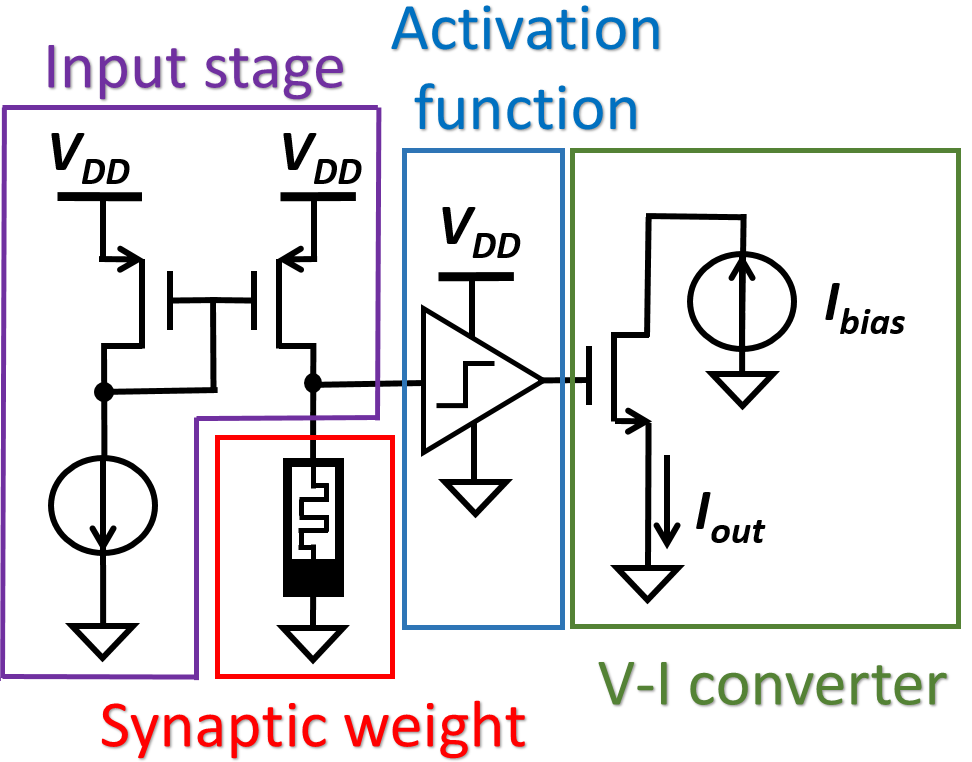}
\caption{Design of single synapse circuit for the HTM architecture \cite{ibrayevdesign}.}
\label{single_synapse}
\end{figure}

The circuit design for the single synapse is illustrated on the Fig. \ref{single_synapse}.
The input stage of the circuit is presented by the current mirror. The input in the form of the current pulses allows for reduced impedance as compared to the cases when the voltage input is used \cite{ibrayevdesign}. The central role in the circuit is dedicated to the memristive device. The memristor is used to perform the function of synaptic weight.  
The weight of the synapse is updated during the learning stage of SP, while during the input phase when the current mirror circuitry is active the permanence value is not affected. This is because the amplitude of the input current signal is selected such that it does not produce voltage higher than the threshold of the memristor, which is the minimum voltage value that is required to change the memristor state.

The purpose of the buffer in the circuit is to establish activation function. High output from the buffer indicates that the weighted input that produced it will affect the calculation of the overlap for the given column. This output is supplied to the voltage-to-current converter that is implemented with help of NMOS. The conversion from voltage to current domain makes the summation of the outputs from the individual synapses for calculation of the overlap value for entire column easier. 

\begin{figure}[!ht]
\centering
\includegraphics[width=3.5in]{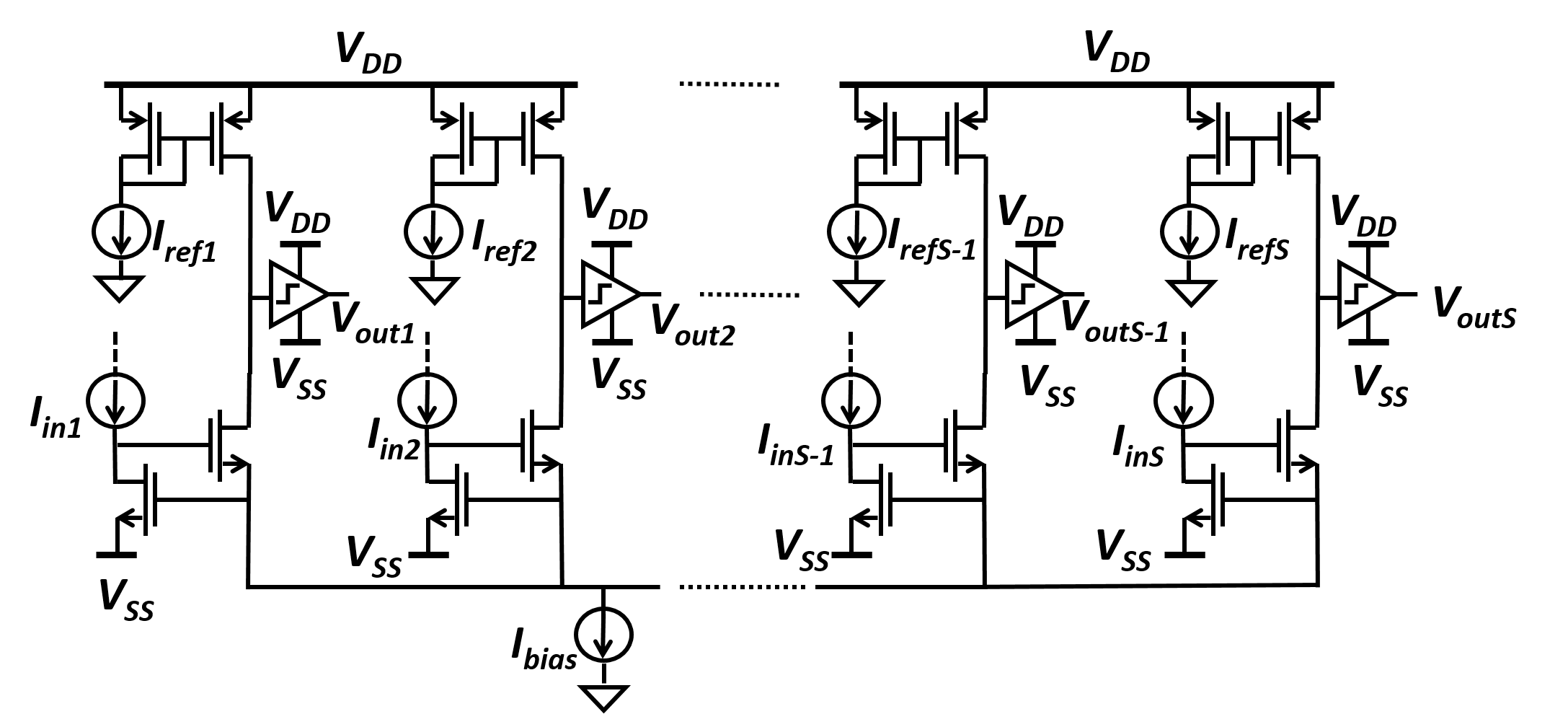}
\caption{WTA circuit \cite{lazzaro1988winner} for realization of inhibition phase of SP.}
\label{wta}
\end{figure}

The inhibition phase of the SP is implemented using WTA circuit that was proposed by Lazzaro et al. \cite{lazzaro1988winner}. Fig. \ref{wta} shows the circuit for the WTA implementation. The inputs ($I_{in1}, ..$$I_{inS}$) to WTA circuit are the total overlap values from different columns. The results from the WTA circuits could be used in the feedback path to update the synaptic weights in the learning phase. 

The proposed design was verified through the face recognition practical problem, where the simulations using AR database resulted in the accuracy of $80 \%$. 
\subsubsection{Memristive array based implementation of SP}
Another design for the SP was presented in the paper of \cite{fedorova2016htm}. The circuit design is based on the memristive crossbar architecture. The validation of the proposed design was performed for the face and speech recognition problems. The proposed systems allows to use SP for extraction of the most important spatial features from the input  image. 
\begin{figure}[!ht]
\centering
\includegraphics[width=3.5in]{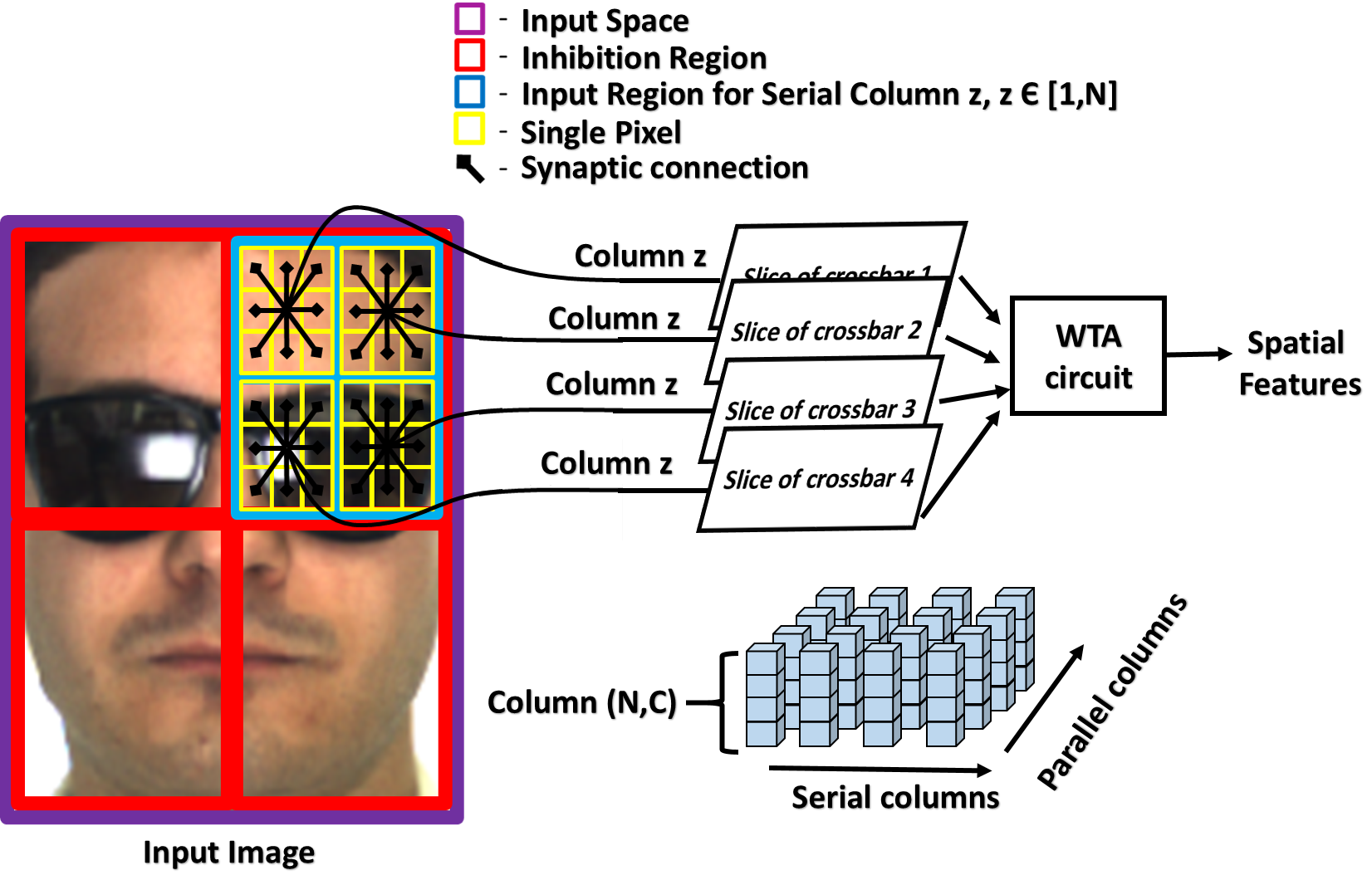}
\caption{ High-level block diagram of the SP implementation \cite{fedorova2016htm}.}
\label{TBCAS}
\end{figure}
The block diagram of the method being used for implementation of the SP by \cite{fedorova2016htm} is presented in Fig. \ref{TBCAS}. The input image is divided into inhibition regions that in turn subdivided into smaller blocks that combine several pixels of the image. Fig. \ref{TBCAS} demonstrates an example of image being divided into 4 inhibition regions with each having 4 small blocks. Single small block in the example contain 9 pixels. During the initialization stage the pixel values are normalized and the weights are randomly distributed across the memristive devices in the crossbars. Several crossbar slices are used in parallel and hence two types of columns are differentiated. Within a single crossbar there is $N$ number of serial columns that are equal to the number of inhibition regions. The number of crossbar slices is related to the number of small blocks inside of   inhibition region, which results in $C$ number of parallel columns. Within a single column there are $S$ number of synapses, which is related to the number of pixels being grouped by a single small block. 

\begin{figure}[!ht]
\centering
\includegraphics[width=3in]{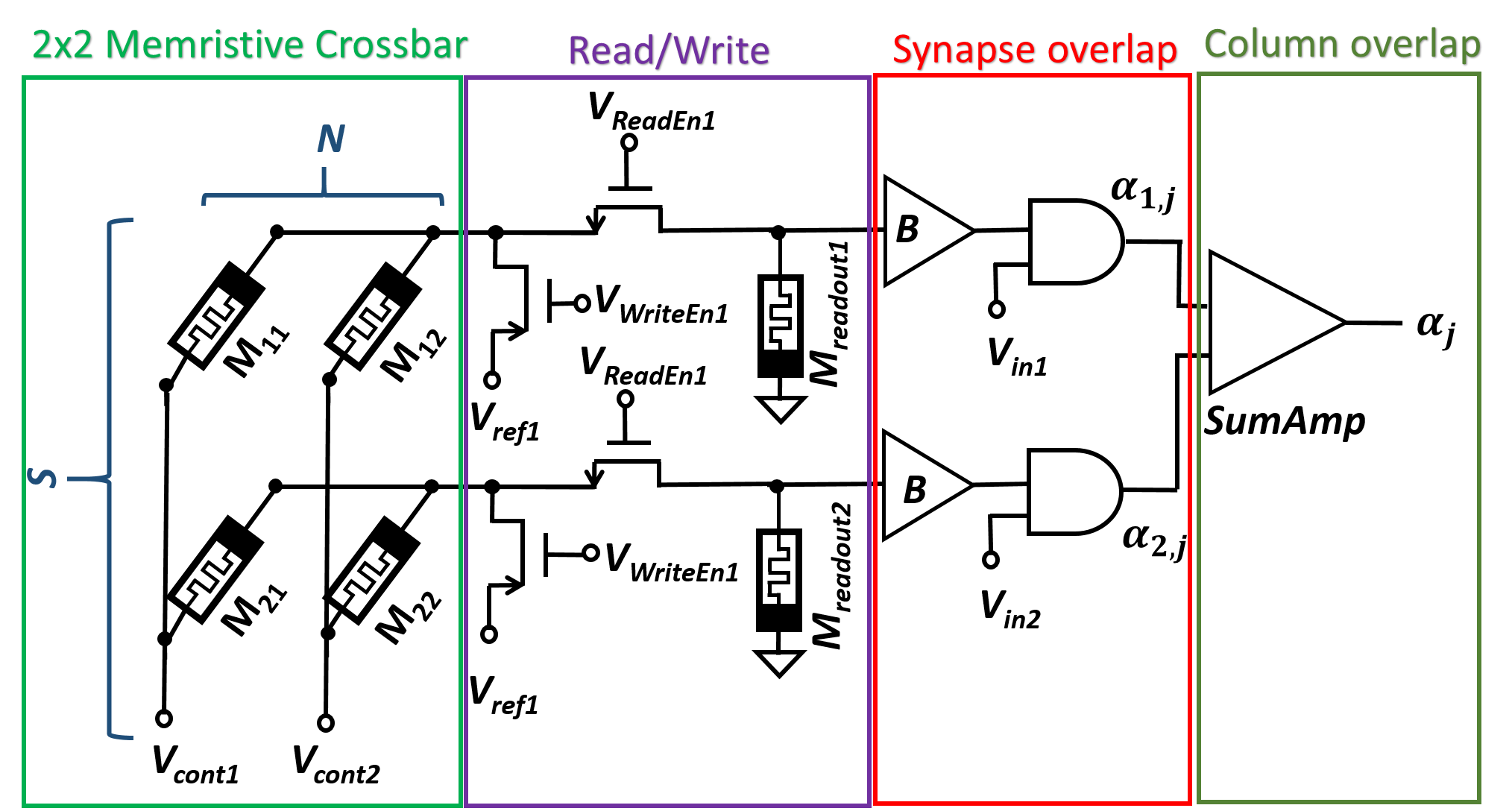}
\caption{Circuit implementation of the memristive crossbar based architecture for overlap calculation stage of SP.}
\label{crossbar}
\end{figure}

Circuit design of the single crossbar slice is presented on the Fig. \ref{crossbar}. It constitutes of memristive crossbar structure, read and write circuitry, circuit for the single synapse overlap calculation and circuit for the calculation of the total overlap of the column. 

{NMOS and PMOS switches allows for transition from read to write mode of the crossbar through the $V_{ReadEn}$ and $V_{WriteEn}$ enabling voltages, respectively. Read and write operations are performed for the single column at a time within a single crossbar slice. { For this to be implemented, the control voltage is applied to the selected column, while all of the other columns are forced to be connected to the ground.
Although this increases the time required for the processing of the single array, consideration of the single column at a time avoids the sneak path problems in the design \cite{yakopcic2013memristor}.}
During the read mode of the crossbars a single serial column within each of the parallel crossbar slices is activated and the voltage is read though the $M_{readout}$ memristor.}
This voltage is used to calculate the overlap values of each of the synapses in the column ($\alpha_{1,j}$, .., $\alpha_{S,j}$). The summing amplifier $SumAmp$ is used to calculate the total overlap value of the column $\alpha_j$. This total overlap values of all of the parallel columns are passed through the WTA circuit \cite{lazzaro1988winner} to determine the winning column. As a result the pixel values that are related to the column that wins among the other columns within the same inhibition region get the value of $1$, while others a set to $0$. This way the important features of an image could be distinguished. 

The output from the proposed SP system is the binarized input image with the features being labeled as important and unimportant. The paper \cite{fedorova2016htm} suggests that the resulted features could be used for the pattern matching and classification applications. For this purpose the 2-bit memristive XOR pattern matcher was realized. The circuit design and explanations for the pattern recognizer will be presented together with the paper on memristive implementation of SP and TM \cite{tcad}.

\subsubsection{Memristive HTM System}

\begin{figure}
\centering
\includegraphics[width=70mm]{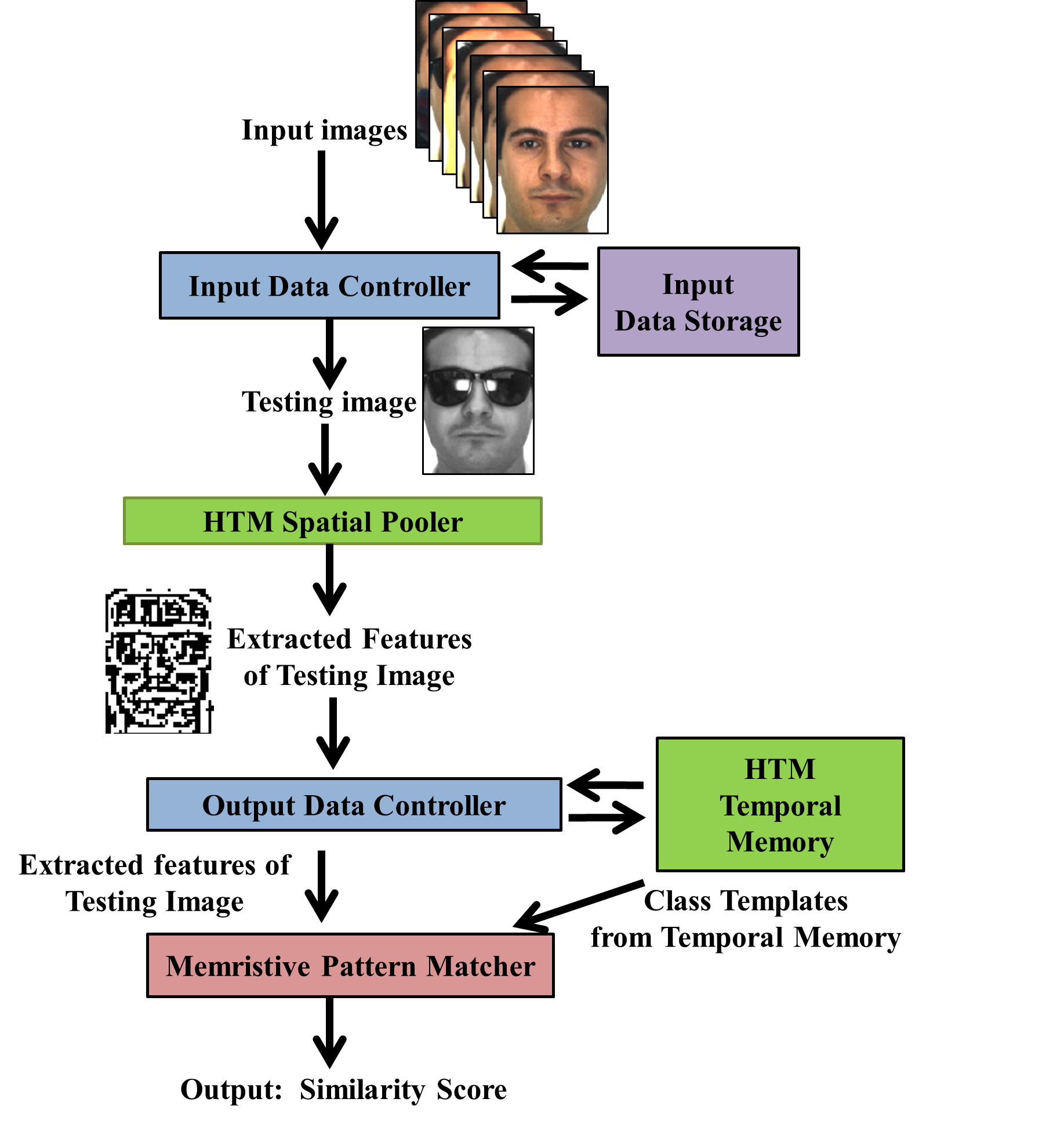}
\caption{The overall system design of HTM face recognition module that consists of the input data controller, HTM SP, HTM TM, output data controller and pattern matcher \cite{ibrayev2017chip}. }
\label{olgafig4}
\end{figure}

\begin{figure}
\centering
\includegraphics[width=70mm]{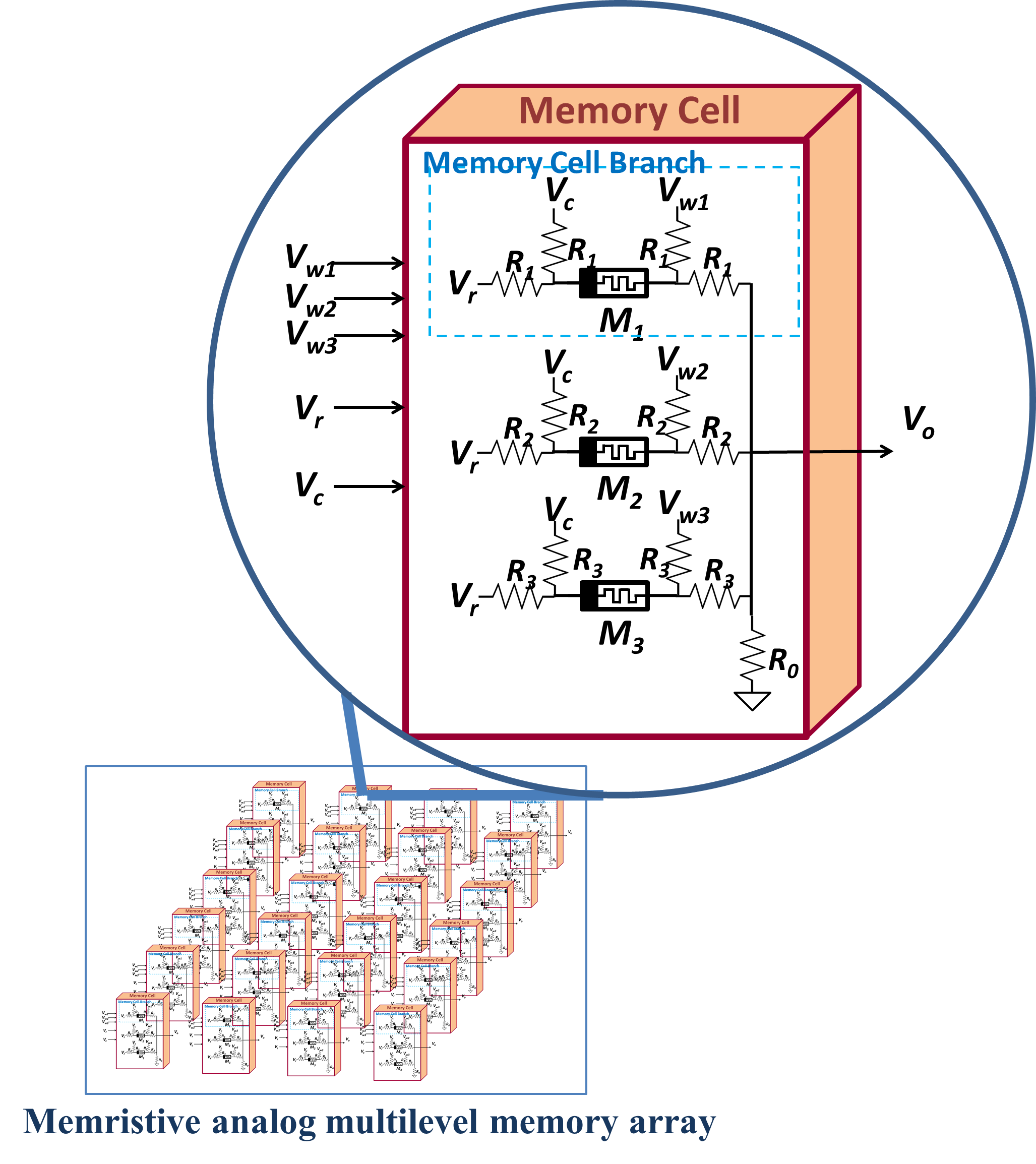}
\caption{Architecture of the analog memristive memory unit involving memristive multilevel discrete memory cell \cite{james2017design}. }
\label{olgafig5}
\end{figure}

The analog implementation of the overall HTM system for face recognition is proposed in \cite{ibrayev2017chip}. The system involves the input data controller, HTM SP, HTM TM, output data controller and pattern matcher. The overall architecture is shown in Fig. \ref{olgafig4}. The input data controller performs the pre-processing of the input data and selects the data samples. The HTM SP is used for the feature extraction. The HTM SP hardware are the same as in \cite{fedorova2016htm}. The HTM SP is followed by HTM TM processing, which is performed on software. However, the HTM TM analog outputs are saved using memristive multilevel analog memory unit shown in Fig. \ref{olgafig5}. The overall architecture of the applied  memristive memory array with multilevel discrete analog memory cells is represented in \cite{james2017design}. The memory cell architecture can consists of 3 or 4 branches with programmable memristors. By applying different write voltage to $V_W$ node, each memristor can be programmed to a particular state. The values of the resistors are different in each branch: $R_1\neq R_2\neq R_3$. In this case, it is possible to store up to 256 analog values with the average error less than 10 \% and up to 1024 values with the average error of approximately 20 \%.  The circuit involved HP $TiO_2$ memristor and the simulations with the Pickett models proposed in \cite{5937942} were modified for large-scale simulations \cite{biol, ascoli2015art}.

\subsubsection{Memristor Logic Circuit for Analog SP and Analog TM}

\begin{figure}
\centering
\includegraphics[width=70mm]{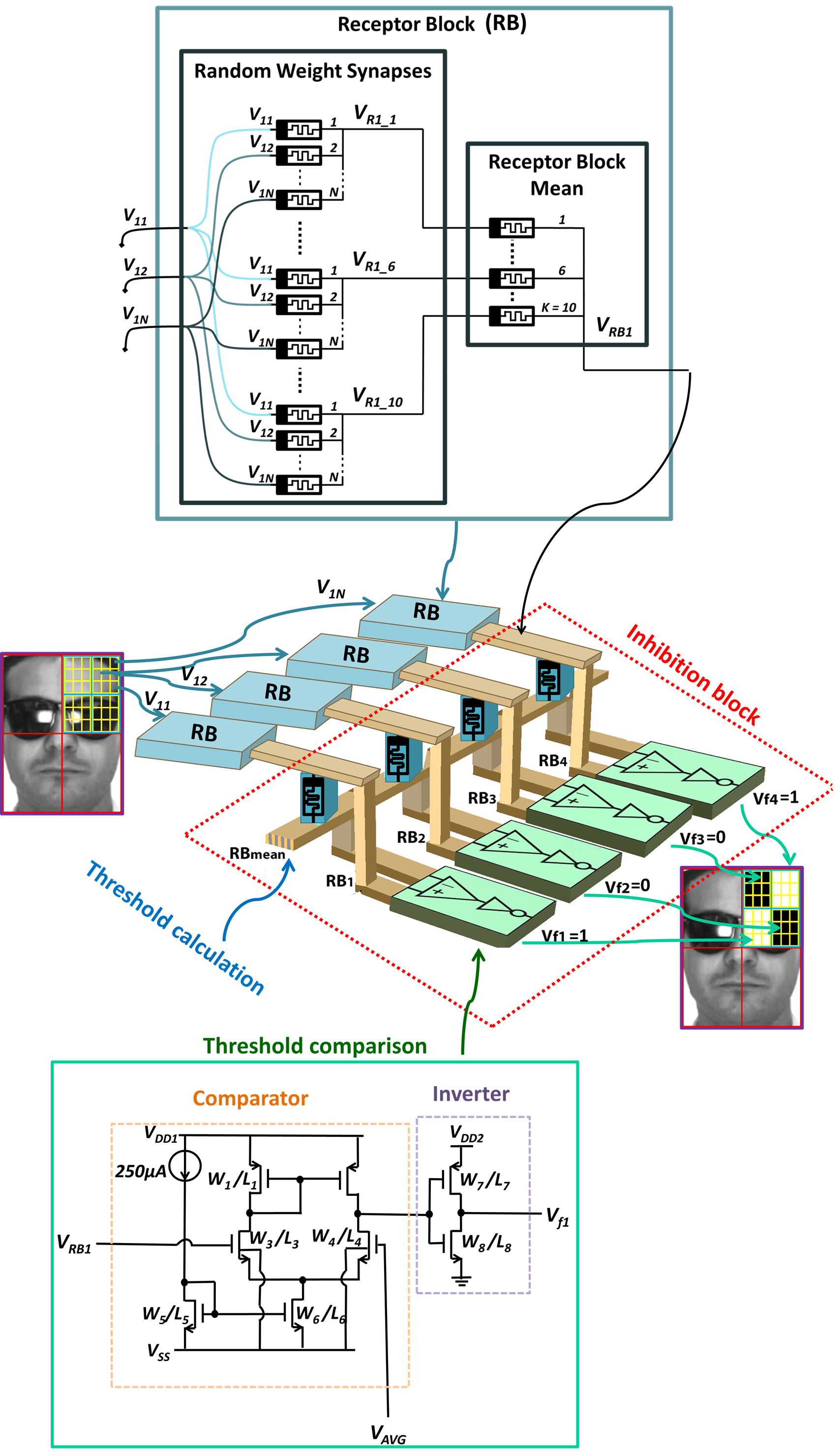}
\caption{Modified implementation of the analog HTM SP with the mean calculation in the inhibition block. Comparing to the other {circuit level implementations} of the HTM SP. This implementation is the closest to the original algorithm and the most efficient in terms of on-chip area and power consumption due to the small size of low-power memristive devices \cite{tcad}. }
\label{olgafig6}
\end{figure}

Recent research illustrated the modified implementation of the analog HTM SP and analog hardware design of the HTM TM \cite{tcad}. Comparing to the previous works \cite{ibrayevdesign, fedorova2016htm, ibrayev2017chip}, the modified HTM SP {analog circuit level implementation} is more scalable and introduces randomization process, which is more accurate and closer to the original HTM SP algorithm. The overall modified HTM SP architecture is illustrated in Fig. \ref{olgafig6}. The main modification of the HTM SP circuit is that it involves the calculation of the mean value in inhibition block instead of the traditional maximum. This brings the significant difference to the {circuit level implementation} of the HTM SP. In comparison to the HTM implementation in \cite{fedorova2016htm}, the WTA and summation stage in the HTM SP is not required and replaced by the memristive averaging circuit. Also, the complete randomization of the HTM SP input weight is added. To perform the randomization phase, the set of averaging circuits with random initial weights are included to the receptor block processing in Fig. \ref{olgafig6}. The weights represented by the resistance of the memristors refer to the synaptic connection. The synapse is either connected, which refers to the $R_{on}$, or disconnected, which is represented by memristors with $R_{off}$. The receptor block output $RB$ represents the calculation of the overlap value of the HTM SP column. 

The threshold calculation block performs the calculation of the average value of all overlaps, which represents the threshold for accepting important features and rejection of irrelevant features. This selection is performed by the inhibition block involving the threshold comparison circuit consisting of the comparator and inverter. If the receptor block output is greater than the threshold value, the output of the inhibition block corresponding to this receptor block becomes $1$, otherwise inhibition block output is $0$. This allows to perform feature encoding with the HTM SP using analog memristive circuits.

\begin{figure}
\centering
\includegraphics[width=80mm]{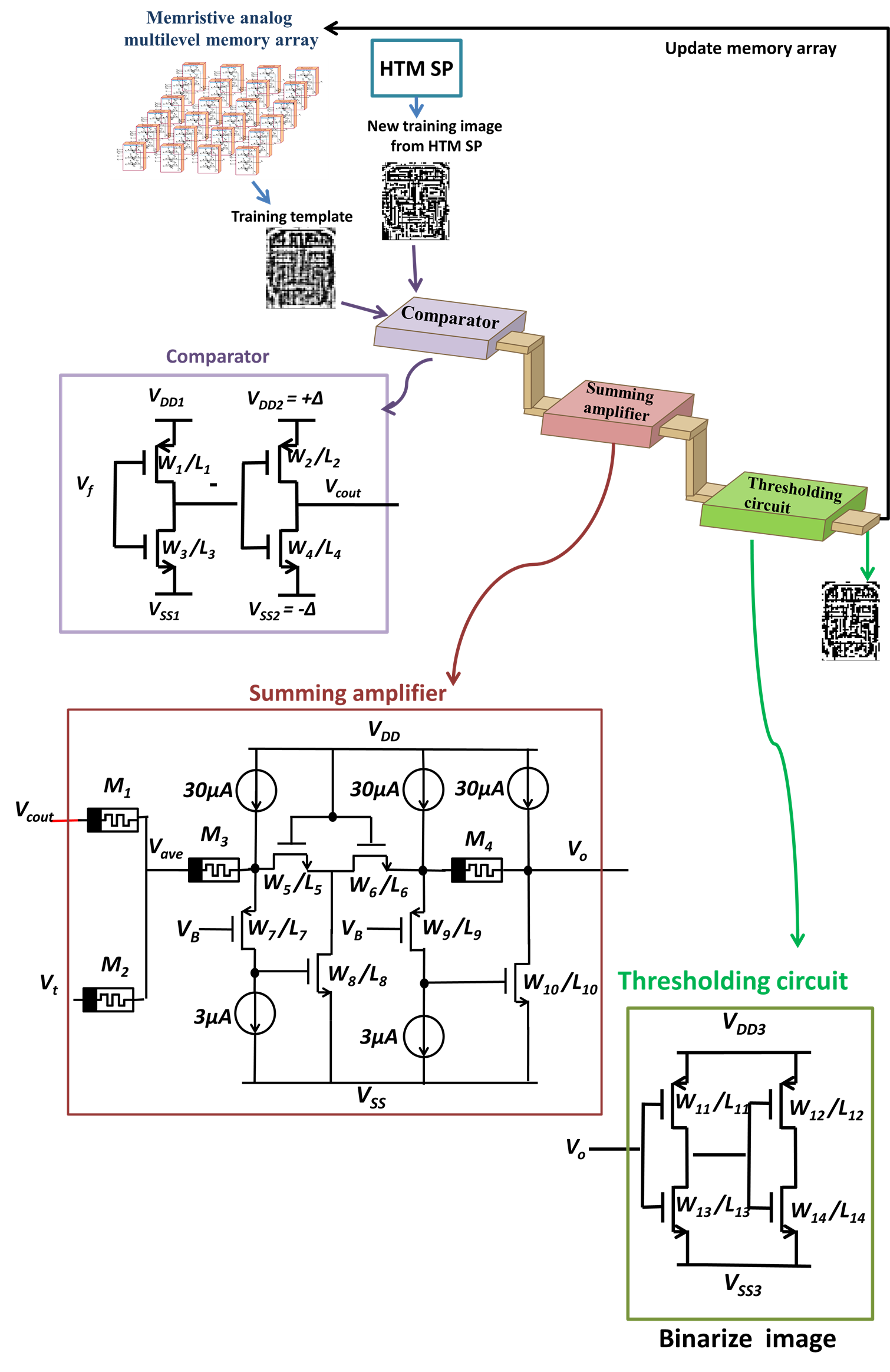}
\caption{The analog {circuit level implementation} of the HTM TM involving comparator, summing amplifier, thresholding circuit and discrete analog memristive memory array. The presented circuit is equivalent to the HTM TM processing of a single input  \cite{tcad}. }
\label{olgafig7}
\end{figure}

The analog {circuit level implementation} of the HTM TM circuits for a single pixel is illustrated in Fig. \ref{olgafig7}. The HTM TM circuit consists of the comparator, summing amplifier, thresholding circuit and discrete analog memristive memory array. 
The class template is stored in the memristive analog memory array introduced in \cite{james2017design}. During the training stage, the data from the HTM SP is fetched to the comparator consisting of two inverters. The second inverter has an underdrive voltage $V_{DD}$, where $V_{DD}=\rho ^+$ and $V_{SS}= \rho ^-$. For the comparator inputs $0$ or $1$, the comparator outputs are $\rho ^-$ and $\rho ^+$, respectively. Then, this comparator output is summed up with the stored template of the particular class using summing amplifier circuit. The summing amplifier consists of the averaging circuit and amplification stage. The averaging circuit calculates the average value between the comparator output and the stored patters; while, the amplification circuit is used to double the average circuit results. This is equivalent to the sum of two inputs. After this, the memory array is updated and obtained results from the summing amplifier is stored. When the training phase is finished, the thresholding circuit, equivalent to the thresholding circuit in the modified HTM SP, is used to binarize the output of the HTM TM. This final binary pattern represents the final result after the training stage corresponding to a particular class, which is further used in the inference stage to perform recognition process.

\begin{figure}
\centering
\includegraphics[width=80mm]{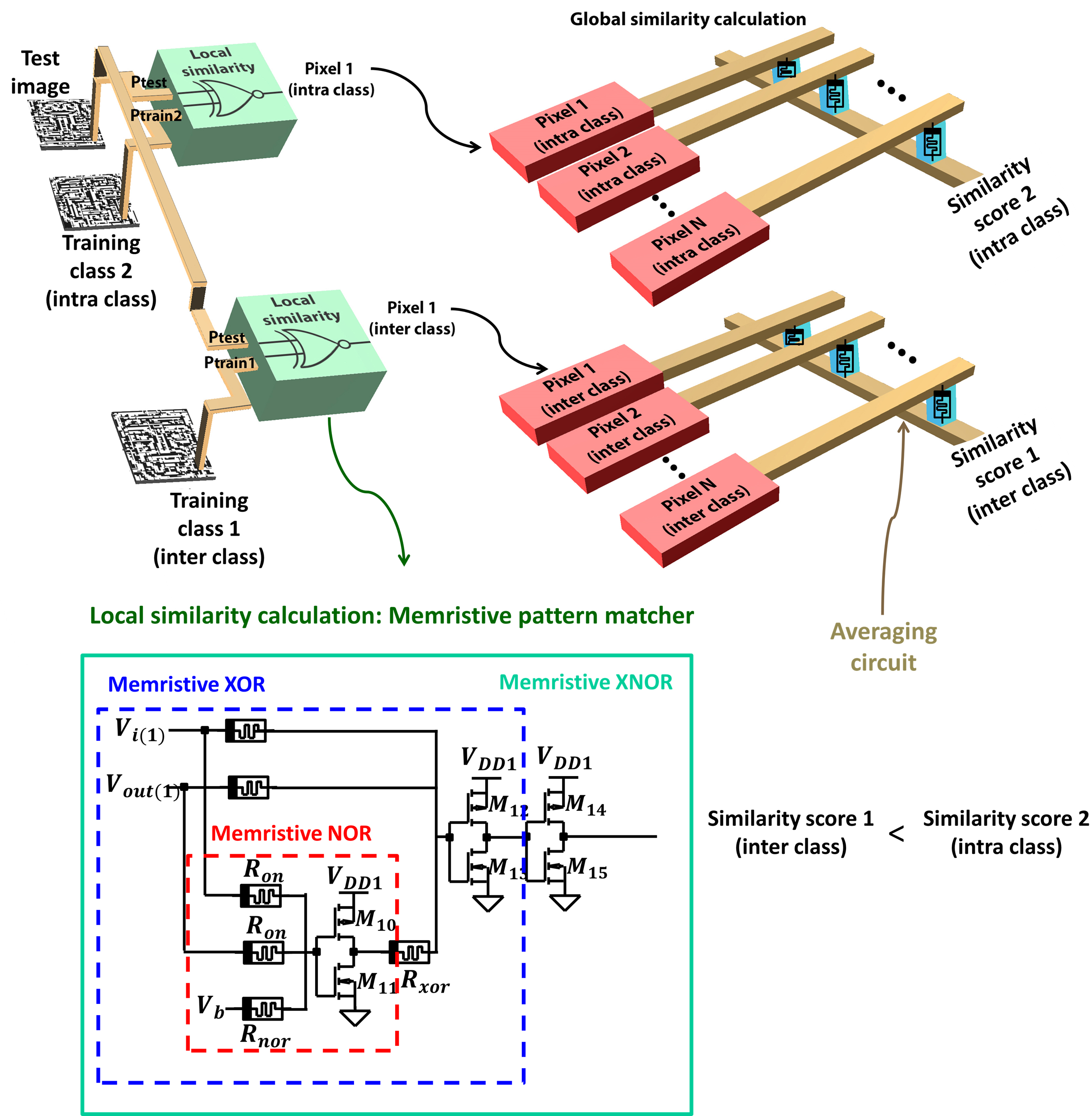}
\caption{The overall pattern classification process involving memristive pattern matcher. The memristive pattern matcher circuit consists of memristive XNOR gates combined with the averaging circuit. The memristive patter matcher is used to obtain the similarity score for a single pixel. Then, the final similarity score for the class is by averaging the similarity scores for all the pixels. The same is performed for different classes. Finally, the output similarity scores for different classes are compared by the WTA \cite{tcad}. }
\label{olgafig8}
\end{figure}

The recognition process in this HTM-based system involves the memristive pattern matcher circuit shown in Fig. \ref{olgafig8}.
The matching circuit is based on the XNOR threshold logic gates \cite{MRTL}. The memristive XNOR gate consists of the memristive XOR gate and inverter. The memristive XOR gate includes the memristive NOR gate comprising three parallel memristors, where two of the memristors are connected to the two input signals that are required to be compared and the third memristor is connected to the control signal $V_c$. The variation of the control signal allows to control the variation of the NOR gate threshold. The XOR gate involves the memristor $R_{XOR}$ that controls the performance and threshold level of the gate. The XOR gate output is inverted to obtain the XNOR gate output. Finally, the output of all memristive XNOR gates for the input pattern are averaged with memristive averaging circuit and the final similarity  score. In the inference stage, the same is performed for the different pattern class samples to compare the current input pattern with all pattern classes obtained after the HTM TM processing. Finally, the pattern matcher outputs of all classes are compared using the WTA circuit.

The extension of the modified HTM SP approach from \cite{tcad} is proposed in \cite{georgia}. The traditional HTM SP  random weight distribution approach is compared to the rule based weight assignment. In this method, the initial weights for the HTM SP are selected based on the local mean intensity of the input pattern within a particular neighborhood. The system level simulation results show that this approach allows to preserve feature sparsity and increase for recognition accuracy for face recognition problems. {The hardware implementation of \cite{georgia} is shown in \cite{krestinskaya2018feature}. Comparing to the HTM SP implementation in \cite{tcad}, modified HTM SP receptor block involves memristive mean calculation circuit, CMOS comparator and CMOS analog switch.} 

{
\subsection{Weight update process}
In all HTM implementations \cite{tcad,georgia,james2017design,fedorova2016htm,ibrayevdesign,ibrayev2017chip}, the HTM weights are programmed as particular resistive states of the memristor. As the memristor is a device that can exhibit various resistive states between $R_{ON}$ and $R_{OFF}$ and possesses a non-volatility property, it can be programmed applying a pulse of a certain duration and amplitude across the memristor. The amplitude of the pulse should be greater than it's threshold voltage.
The duration and amplitude of the pulse depends on the memristor characteristics, such as switching threshold, set time, reset time, $R_{ON}$ and $R_{OFF}$ parameters. The parameters vary based on the memristor material and size of the device. The number of possible resistive states depends on the memristor properties \cite{vourkas2016emerging}. 
In the architectures with single synapses that are not connected to each other, the memristors are easier and faster, while in a crossbar architectures and accurate switching between crossbar rows and columns is required and the update process is slower.}

{
\subsection{Comparison of the memristive HTM designs}
\label{snew}
Each memristive HTM design has particular advantages and drawbacks. The mixed signal design presented in \cite{fan2016hierarchical} calculates a dot product in HTM SP and HTM TM using memristive crossbar, where less CMOS components are required and total area and power consumption are reduced, comparing to the analog HTM implementations and equivalent FPGA implementation \cite{fan2016hierarchical}. The efficient method of dividing a large image into smaller crossbars is applied, which allows to avoid common scalability issues related to the crossbar implementation.
However, in this approach the training of weights is done offline on the software,
and only inference phase is performed on hardware, which make it impossible to be used for near-sensor edge computing applications. In addition, as the emerging spin-neuron devices are used, the compatibility with CMOS/memristive circuits should be investigated further.

The memristive HTM design in \cite{ibrayevdesign} represents only a single synapse circuit for HTM cell and it is one of the earliest works on circuit level implementation of HTM. The synapse has many CMOS components; thus it is not feasible to scale such circuit for large scale simulations. The circuit might be appropriate for the applications with small number of inputs, however for image related application, on-chip area and power consumption of such circuit is large. In addition, the accuracy of system level implementation of equivalent algorithm is low.

The HTM design in \cite{fedorova2016htm} is based on the memristive crossbar that can be scaled comparing to \cite{ibrayevdesign}. In addition, the overall HTM accuracy is higher. However, to achieve high accuracy, the image pre-processing and filtering stage is required. Even thought the online training is proposed, the training process of large memristive crossbar is slow, because the programming of each memristor to a certain weight is required and a single column with memristors is programmed at a time.

One of the advantages of HTM designs in \cite{tcad} is the proposed implementation of HTM TM circuit with online learning. In addition, the training process in HTM SP is performed faster, comparing to the crossbar implementation because of the use of separate memristors and the possibility to program them in parallel. However, the design contains more CMOS components, therefore on-chip area and power consumption are large.

The HTM approach represented in \cite{georgia} and \cite{krestinskaya2018feature} explores the possibility to remove a learning circuit from the HTM SP. Even though the area and power of receptor block is increased, the absence of the HTM SP learning phase compensates this. The accuracy for particular application and database is high.
However, this approach is appropriate for a limited set of image processing applications and the hardware implementation is not fully investigated.
}

\section{Applications of Memristive HTM}
\label{s4}

\subsection{Overall approach for system level simulation of HTM}

\begin{figure}
\centering
\includegraphics[width=90mm]{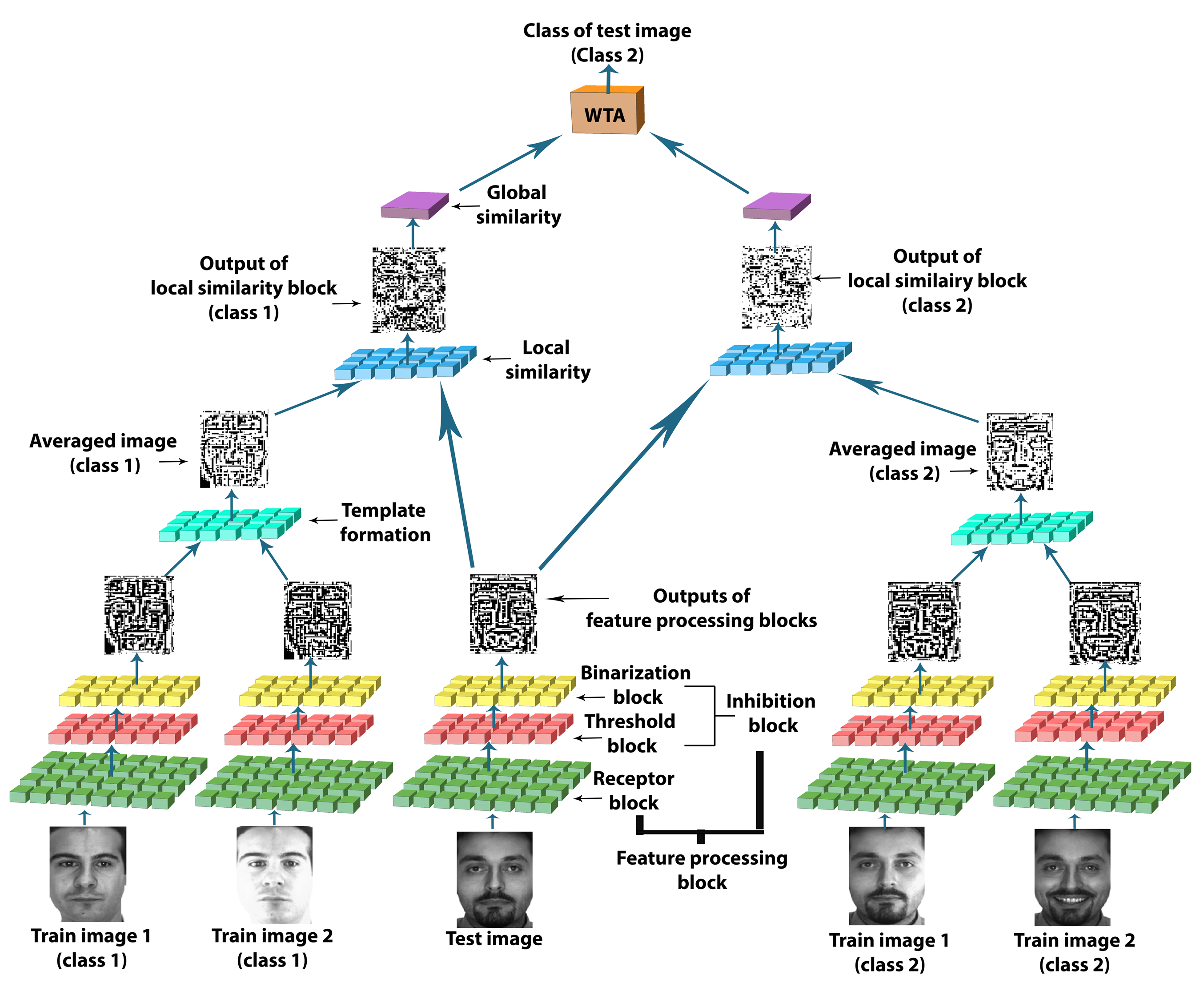}
\caption{Example of the system level processing using HTM for the face recognition problem. The system level implementation of HTM includes the HTM SP, HTM TM and inference stage involving calculation of local and global similarity scores and comparing them using the WTA algorithm. }
\label{olgafig9}
\end{figure}

\begin{figure}
\centering
\includegraphics[width=65mm]{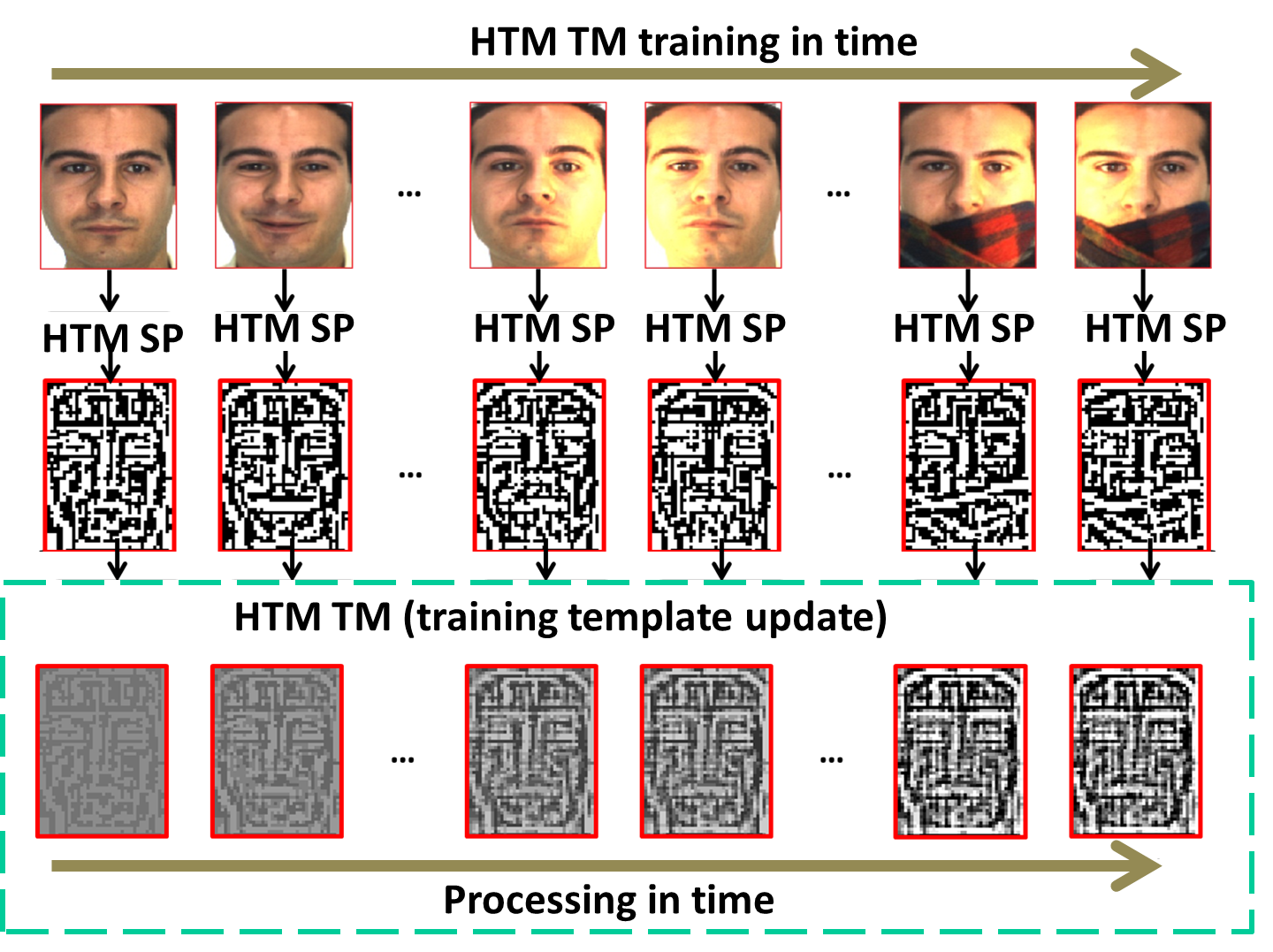}
\caption{Example of the HTM TM processing and change and update of the class template with time \cite{tcad, ibrayev2017chip}. }
\label{olgafig10}
\end{figure}

The example of HTM system level processing for face recognition application is shown in Fig. \ref{olgafig9}. The pre-processing phase including filtering and image normalization is not included to this example. The images are processed in a hierarchal manner starting from the HTM SP processing involving the image block processing and inhibition region processing. After the HTM SP, the sparse binary representation of the input image is formed. Fig. \ref{olgafig9} shows the example of the processing of testing image and comparison of it to 2 training classes. 
{In the training phase, the training image template is formed by the HTM TM combining all intra-class training patterns. The training phase includes the update of the weights of HTM synapses using Hebbian learning rules and backpropagation algorithm to minimize the recognition error.
Next, in the testing phase, these trained weights are used.} The test image is compared with the training templates of all classes and the similarity score for each class refering to each feature in the image (called local similarity) is calculated. The global similarity score for each class is calculated combining all local similarity scores of the same class by either summation or averaging operation. Finally, the global similarity scores are compared by the WTA function that selects the winning class. Fig. \ref{olgafig9} illustrated the example of 2 training image in the HTM TM. If the number of training images is larger, the class template representing a particular pattern class is updated in time. Fig. \ref{olgafig10} illustrates this process. Each training image is initially processed by the HTM SP before fetching to the HTM TM. The class template evolves with time during the training process with the HTM TM.

{ 
The parameters of the HTM SP and HTM TM vary depending on the application. Such parameters as number of cells, layers, columns per HTM region are selected experimentally and adjusted according to the application. For example, the parameters of HTM for NPL related problems \cite{poster} are different from the image classification problems \cite{tcad,fedorova2016htm}. In research works \cite{tcad}, \cite{ibrayev2017chip} and \cite{fedorova2016htm}, the authors select the number of columns and number of regions experimentally performing system level simulations in Matlab simulations and verifying the best possible accuracy for particular applications and databases. In addition, even for the same application the number of columns can vary for different databases, and the optimum parameter for particular application can be selected based on experimental verification of the maximum possible accuracy for all databases \cite{tcad}. Depending of the number of cells, layers and columns in HTM, the complexity of the circuits varies. For example, in modified HTM \cite{tcad}, the number of regions in HTM effects the total power consumption and on-chip area, while the number of columns effects only on-chip area. In \cite{tcad}, the trade-off between the accuracy variation with the number of columns and corresponding on-chip area and power consumption for particular image classification application is discussed. 
}

\begin{algorithm}
\caption{System level implementation of HTM}\label{alg:matlab}
\begin{algorithmic}[1]

\State $x = grayscale(x)$
\Comment{\textbf{PRE-PROCESSING}}
\State $x = filtering(x)$
\For {$p$ inhibition regions}
\Comment{\textbf{HTM SP}}
\For {$k$ image blocks} 
\For {$j$ iterations} 
\State $W = random(S,S)$
 \ForAll{$i \in W$}
 \If{$W(i) \geqslant \theta$} 
 \State $W(i) = 1$
 \Else
 \State $W(i) = 0$
 \EndIf
 \EndFor
\State $image.block (j) =mean(W(j)\times image.block(j))$ 
\EndFor
\State$image.block(k)=mean(all.image.blocks (j))$
 \EndFor
\State$threshold.block = mean (y.image.blocks)$
\For{$y$ image blocks}
\If{$image.block(y) > threshold.block$} 
\State $inhibition.region(y) = 1$
\Else
\State $inhibition.region(y) = 0$
\EndIf
\EndFor
\State $x(p) = inhibition.region(p)$
\EndFor
\If{training phase}
\Comment{\textbf{HTM TM}}
\For {$c$ training classes}
\For{$t$ training images}
\For{all $j$ pixels in $x$}
\If {$x(j)$ = 1}
\State$\begin{aligned}[]
             class.t(c,j)=class.t(c,j) +\rho ^+
         \end{aligned}$
\ElsIf {$x(j)$ = 0}
\State$\begin{aligned}[]
             class.t(c,j)=class.t(c,j) +\rho ^-
         \end{aligned}$
\EndIf
\EndFor

\For{all $j$ pixels in $x$}
\If {$class.t(c,j) > \gamma$}
\State$class.t(c,j) = 1$
\ElsIf {$class.t(c,j) < \gamma$}
\State$class.t(c,j) = 0$
\EndIf
\EndFor
\EndFor
\EndFor
\ElsIf{testing phase}
\Comment{\textbf{RECOGNITION}}
\EndIf
\For{$c$ classes}
\For{$j$ image pixels}
\State$score(c)=sum(XOR(class.t(c,j),x(j)))$
\EndFor
\EndFor
\State$class=class(min(score))$
\end{algorithmic}
\end{algorithm}

The system level algorithm for the HTM system implementation is shown in Algorithm \ref{alg:matlab}. The system is divided into four phases: pre-processing, HTM SP, HTM TM and recognition. The pre-processing phase, corresponds to lines 1-2 in Algorithm \ref{alg:matlab} and usually is performed by the input data controller or involves additional software based processing. In the pre-processing phase, the input patterns $x$ are converted to gray scale, normalized and filtered. In \cite{tcad,fedorova2016htm,ibrayev2017chip}, the application of the standard deviation filter is proposed for the image pre-processing. According to \cite{tcad, georgia}, the filtering process can replace the learning phase of the HTM SP and increase the recognition accuracy in pattern recognition problems.

HTM SP processing refers to lines 3-20. 
First, the input pattern (image) is divided into $p$ inhibition regions and each inhibition region contains $k$ image blocks. For $j$ iterations, the image block calculation is performed and the final image block is calculated as a mean of all image blocks obtained after $j$ iterations. To calculate the image block value for a single iteration, the $S\times S$ random weight matrix $W$ with values between 0 and 1 is generated and the matrix weights are binarized based on the threshold value $\theta$. The image block for a single iteration is calculated by averaging of the product of weights $W(j)$ and image block ($j$). The final value of a single image block is obtained by averaging of all calculated image blocks from $j$ iterations.
The $threshold.block$ is formed as an average value of $y$ image blocks. Each threshold corresponds to a single inhibition region. The inhibition region output is obtained by comparing the inhibition region to the threshold block. If the value of the image block output is greater than the corresponding threshold block value, the $inhibition.region(y)$ is calculated as 1; otherwise, $inhibition.region(y)$ is 0. The output from the HTM SP is formed by concatenating $p$ inhibition regions.

After the HTM SP, either the training phase with HTM TM or the inference phase involving pattern classification is performed. The HTM TM processing corresponds to the training phase and refers to lines 21-33. The value $c$ represents the number of possible classes, $t$ is the number of training images, and $j$ is the number of pixels in a single training image. In the HTM TM, the class template $class.t$ is updated with each training pattern. 
 The initial $class.t(c)$ value corresponds to the first training pattern of the class $c$. The update process involves the traditional Hebbian rule. Based on the pixel values of new training images, the $class.t$ is updated by adding  $\rho ^+$ or $\rho ^-$ value to each feature (pixel) of the training class template $class.t$. After the last training pattern, the training template is binarized based on the threshold $\gamma$.

{
The inference phase varies with the application of HTM. For prediction making applications, HTM TM is involved in inference phase and the predictions are made based on the learned training sequences. For the pattern recognition applications, inference phase involves only recognition and classification process shown in lines 34-38.} All features of the input image $x(j)$ are compared with the corresponding pixels of the class templates $class.t(j)$ of each class $c$ using the $XOR$ function. Then, the similarity score $score(c)$ for each class $c$ is calculated. Finally, the class is determined by minimum similarity score corresponding to the minimum distance to the class template.

\subsection{HTM Applications}
The applications of the HTM algorithm and hardware implementation include face, speech and gender recognition, natural language processing (NLP) and recognition of handwritten digits. Table \ref{table} presents a comprehensive comparison analysis of the different  hardware implementations of HTM that were discussed in the survey. The designs are compared in terms of technology being used for the implementation, resulted recognition/classification accuracy, area and power requirements and practical application to which proposed system could be applied.

\subsubsection{Face Recognition}
Most of the analog HTM implementations focus on the face recognition problems. The examples of the HTM application for face recognition are shown in \cite{ibrayevdesign, fedorova2016htm, tcad, ibrayev2017chip,james2017design, georgia}. The works involve the application of AR \cite{martinez1998ar}, Yale \cite{47} and ORL databases \cite{46}. The HTM processing face images usually involves the conversion of the $RGB$ images into gray scale and filtering using standard deviation filter. The possible accuracy of the recognition varies from approximately 63\% up to 98\% for different configurations of HTM.

\subsubsection{Speech Recognition}
The speech recognition involving HTM is represented in \cite{james2017design, tcad, fedorova2016htm}. The speech recognition is performed using TIMIT database \cite{garofolo1993darpa}. The speech sequences are transfered into the pattern matrices and then processed as the images. The maximum speech recognition accuracy that can be achieved using HTM is approximately 95\%.

\subsubsection{Handwritten Digits Recognition}
The digital and mixed-signal HTM implementations focus mostly on handwritten digits recognition represented in \cite{streat2016non, fan2016hierarchical}. The maximum recognition accuracy for the digital HTM implementation for handwritten digits recognition is 95\% and 92\% for mixed-signal and digital implementation, respectively.

\subsubsection{Gender Classification}
The HTM was also tested for gender classification problem \cite{james2017design}. The gender classification is similar to the face recognition and performed using \cite{lenc2015unconstrained}. For the gender recognition the HTM SP with TM performed better than the single HTM SP. The average accuracy achieved for the gender classification is approximately 71\% and 63 \%, respectively.

\subsubsection{HTM for Natural Language Processing}
The recently proposed HTM application for NLP is represented in \cite{poster}. It involves the sequence learning for detecting patterns of any natural language that is represented as a system containing several levels of symbols. It has been proposed to apply HTM for correction misspelled words on hardware level. However, the simulation results or {circuit level implementation} of the HTM for this purpose has not been presented yet.

\begin{table*}[!ht]
\centering
\caption{Comparison of the hardware implementations of HTM}
\label{table}
\begin{tabular}{|p{0.63cm}|p{2.9cm}|p{2.9cm}|p{2.3cm}|p{3.25cm}|p{1.8cm}|p{1.4cm}|}
\hline
\textbf{Papers} & \textbf{Specifications} & \textbf{Technology}                                                                          & \textbf{Application} & \textbf{Accuracy} & \textbf{Area} & \textbf{Power} \\ \hline
\cite{fan2016hierarchical}    &      \begin{tabular}[c]{@{}l@{}} Memristive crossbar, \\ HTM SP and TM\\   \end{tabular}       &   \begin{tabular}[c]{@{}l@{}}Mixed-signal design:\\ $Ag$-$Si$ memristors \\ and spin-neuron devices\end{tabular}                                                                                      &   \begin{tabular}[c]{@{}l@{}} Handwritten\\ digits recognition  \end{tabular}                  &                    up to 95 \%   &        NA   &  NA           \\ \hline
\textbf{\cite{ibrayevdesign}}  &   \begin{tabular}[c]{@{}l@{}}  Memristor-CMOS\\ hybrid design, \\ HTM SP \   \end{tabular}                  & \begin{tabular}[c]{@{}l@{}}Analog design:\\ 90nm IBM CMOS, \\ Yakophic memritor\\ model \cite{yakopcic2013memristor}\end{tabular} &     Face  recognition                 &    80\% (AR database)             &   \begin{tabular}[c]{@{}l@{}}56.19$\mu m^2$ \\ (9x9 array)     \end{tabular}        &         4.79$\mu W$       \\ \hline
\textbf{\cite{fedorova2016htm}}          &    \begin{tabular}[c]{@{}l@{}}Memristive crossbar, \\ HTM SP\end{tabular}                  & \begin{tabular}[c]{@{}l@{}}Analog design:\\ 180nm IBM CMOS,\\ Biolek memritor \\model \cite{biol}\end{tabular}    &  \begin{tabular}[c]{@{}l@{}} Face recognition\\ Speech recognition \end{tabular}                   &   86.42\% (AR database), 86.67\% (Yale database),  70\% (Average for speech)               &      \begin{tabular}[c]{@{}l@{}}   12.51$\mu m^2$ \\ (9x9 array)   \end{tabular}  &    31.56$\mu W$     \\ \hline

\cite{ibrayev2017chip}         &  \begin{tabular}[c]{@{}l@{}}Memristive crossbar \\ with analog memristive \\ memory array , \\ analog HTM SP, \\HTM TM on software\end{tabular}                      &           Analog design                                                                                   &       \begin{tabular}[c]{@{}l@{}}Face recognition \end{tabular}              &         \begin{tabular}[c]{@{}l@{}}HTM SP: 76.54\%,\\ HTM SP and TM: 83.48\% \end{tabular}            &    NA           &  NA              \\ \hline

\cite{james2017design}          &   \begin{tabular}[c]{@{}l@{}}HTM with analog\\ memristive memory array,\\analog HTM SP, \\HTM TM on software\end{tabular}                      &           \begin{tabular}[c]{@{}l@{}}Analog design:\\ 180nm IBM, \\Modified Pickett \\model \cite{abdalla2011spice}\end{tabular}    &  \begin{tabular}[c]{@{}l@{}}Face recognition ,\\ \\Gender Classification\\  \\Speech recognition\end{tabular}              &         \begin{tabular}[c]{@{}l@{}}HTM SP: 76.54\%, \\HTM SP and TM: 83.48\%, \\HTM SP:63.06\%, \\HTM SP and TM: 71.17\%,  \\HTM SP:up to 70\%,\\ HTM SP and TM: up to 90\%\end{tabular}            &   NA           &         NA      \\ \hline

\cite{tcad}           &   \begin{tabular}[c]{@{}l@{}} Memristive-CMOS design \\ involving analog memory \\ and memristive pattern \\ matcher,\\ analog HTM SP and TM \end{tabular}               &     \begin{tabular}[c]{@{}l@{}}Analog design:\\ 180nm {TSMC}, \\Biolek memritor \\model \cite{biol}\end{tabular}                                                                                             &        \begin{tabular}[c]{@{}l@{}}Face recognition, \\Speech recognition \end{tabular}                                   &   \begin{tabular}[c]{@{}l@{}}  87.21 \%  \\up to 95\% \end{tabular}          &    \begin{tabular}[c]{@{}l@{}}HTM SP \\ ($1\times4$):\\$19.96 \mu m^2$, \\HTM TM \\ ($1\times1$): \\$23.85 \mu m^2$ , \\Pattern matcher \\($1\times1$):\\ $1.18 \mu m^2$ \end{tabular}     &     \begin{tabular}[c]{@{}l@{}}HTM SP\\($1\times4$):\\$365.88 \mu W$, \\HTM TM\\ ($1\times1$): \\$442.26 \mu W$, \\Pattern\\matcher \\($1\times1$):\\$69.44 \mu W$ \end{tabular}    \\ \hline

{\cite{georgia}, \cite{krestinskaya2018feature}   }   &  \begin{tabular}[c]{@{}l@{}}Algorithm modification \\of HTM SP \cite{tcad} \\ with hardware testing \cite{krestinskaya2018feature} \end{tabular}       &   \begin{tabular}[c]{@{}l@{}}Analog HTM SP \\modified design of \cite{tcad},\\   modified algorithm in \cite{georgia}, \\hardware testing \cite{krestinskaya2018feature} \end{tabular}                                                                                         &    Face recognition     &    up to 98 \%               &      {Single receptor block: $13.31 \mu m^2$   }     &    {     $135 \mu W$  }     \\ \hline




\end{tabular}
\end{table*}

\section{Discussion}

\label{s5}

\subsection{Advantages of the Memristive Implementation of HTM}

{To discuss the advantages of the memristor based circuit realizations of HTM over the other attempted approaches, a brief overview of the non-memristive digital implementations is required. 
Recent publications on the digital implementation of HTM modules include FPGA platforms \cite{zyarah2015design, streat2016non} and VLSI design \cite{melis}. 
The circuit design for non-volatile architecture for HTM implementation using VHDL  implemented in \cite{streat2016non} requires  $104.26mm^2$ of area footprint 
and $64.394mW$ of power for an 8-channel based SP model and demonstrates classification accuracy of 91.89\% for the MNIST database.
The FPGA implementation of HTM \cite{zyarah2015design} allows to speed up the training process and produce the accuracy of 91\% for MNIST and 90\% for EUNF \cite{padilla2013performance} databases even with the presence of noise. While the accuracy of approximately 95\% can be achieved for prediction making using HTM algorithm \cite{melis}. }

Comparing to the digital implementations of the HTM algorithm \cite{zyarah2015design, streat2016non, melis}, the advantages of the analog memristive HTM implementations {that were presented in this paper} are related to the processing speed, small on-chip area and adjustable power dissipation. The {full analog circuit level implementation} of HTM would allow to process visual data faster, in contrast to the digital implementations that are limited by the processing frequency of FPGAs and the sampling rate of analog-to-digital and digital-to-analog converters. 

The memristive devices used in the HTM design for various components, such as amplifiers, pattern matcher and averaging circuits, ensure the reduction of the on-chip area of the HTM, in comparison to the implementation that are based on resistors or fixed digital FPGAs and VLSI implementations. Also, while the digital HTM implementations have fixed area and power dissipation, the analog circuit for HTM can be adjusted. For example, the high power components, such as operational amplifiers, can be replaced by the circuit components with lower power dissipation. {In addition, the other advantage of using memristors, comparing to the CMOS transistors and resistors, is low leakage currents within the device.} 

\subsection{Limitations of Memristive HTM Hardware Implementation}

The HTM SP algorithm is simplified to implement it using analog hardware. Such parameters as more than $1$ active columns after overlap calculation, update of the boosting factor, complete Hebbian-based learning have not been implemented yet.
In recently proposed studies \cite{ibrayevdesign,tcad,fedorova2016htm, ibrayev2017chip}, the number of winning columns in the overlap phase of HTM SP is limited to $1$ because of the overlap parameter is selected using the traditional WTA circuits. However, in the original algorithm the desired number of winning columns can be selected. To improve this limitation, the implementation of the HTM SP with adjustable number of outputs in the WTA circuit is required.

In addition, in most of the implementations the boost factor parameter is not taken into account or represented as $1$. This is appropriate for the visual data processing applications \cite{ibrayevdesign,tcad,fedorova2016htm, ibrayev2017chip}. However, the original HTM algorithm implies that the boosting factor should be updated \cite{25}. The analog {circuit level implementation} of the boosting factor update may open the possibility of the successful application of HTM not only for the pattern recognition and visual data processing, but also the other applications, such as prediction making.

The existing {circuit level implementation} of the HTM TM proposed in \cite{tcad} is not scalable. The HTM TM circuit is presented for a single pixel implementation. The real time implementation of such hardware circuits involves the consideration of the trade-offs between the parallel and sequential processing of the input data. For the parallel processing of the HTM TM circuits, the on-chip area and power is large and it is difficult to scale the architecture. However, the processing speed of such circuit is high, comparing to the implementation of the equivalent digital circuits or sequential processing of the input data using  the same HTM TM design. The sequential processing requires the additional sequence control circuit. 

\subsection{Open Problems}
 
One of the most important problems in the hardware implementation of HTM is the continuous hardware training, learning and parameter update process. The research study \cite{25} states that the boosting factor in the HTM SP should be updated based on the average-activity of the mini-columns considering $T$ previous inputs. The appropriate number of $T$ is selected to be 1000, which means that 1000 update cycles in analog hardware is required. Considering that the learning and update process in \cite{tcad} is performed using analog memristive memories \cite{james2017design}, the real on-chip implementation of the stable multilevel memristive memory with the long lifetime is one of the challenges in the analog {circuit level implementation} of the memristive HTM. Therefore, the robust {analog circuit level implementation} for online learning and training and of HTM is required. The fully analog learning circuit for the HTM will remove the necessity of the  offline software-based or FPGA-based training process, which would improve the processing speed. Also, the implementation of the efficient and robust analog memory unit is necessary to achieve the maximum processing speed and efficient temporal data storage. The implementation of the on-chip analog memory unit for the memristive HTM will remove the necessity to have the separate storage unit, which will improve the processing speed and reduce required on-chip areas and power consumption.

{The implementation of HTM with backpropagation learning has not been fully investigated in software implementations of HTM, and the circuit implementation of HTM with backpropagation has not been proposed yet. Even though theoretical HTM works mention that the backpropagation with gradient descent is one of the possible ways to update HTM weights \cite{tmnew}, the HTM with backpropagation remains an open problem. The implementation of such circuit can increase the accuracy of the results and decrease the number of iterations in the training stage, because the weight update is much accurate in backpropagation learning comparing to the simple Hebbian rule. The Hebbian learning rule involve a simple update by $\pm \Delta$ value depending on the sign of error, while the backpropagation with gradient decent considers the gradient of error in each update and the updates in each weight are different depending of the amount of error. The analog circuit implementation of such algorithm may lead to the decreased number of required iteration, which insures less endurance issues and longer lifetime of the memristive weights. However, the time required for update process may increase, especially in a crossbar, where each weight has to be updated separately. In addition, the complexity of the circuits, on-chip area and power consumption also increase, because additional CMOS components are required for the implementation.}

To solve the analog HTM scalability problems, the analysis of the parasitics, {sneak currents} and  leakage currents of the large memristive CMOS circuits is required. 
{As in most of the cases gateless \cite{zidan2014leakage} memristive arrays are used, which allows to avoid additional leakage current comparing to the gated arrays with CMOS transistors, the sneak path problems of such arrays for HTM applications should be investigated.}
The scalability of the memristive crossbar array for analog HTM design should be tested and the limitations of the number and side of the HTM input should be identified. Also, the performance of the memristive HTM SP and TM in terms of recognition and classification accuracy should be tested on the real memristive crossbars considering influence of the parasitic effects and sneak currents on the HTM performance.

The realistic implementation of memristive HTM circuits required at least the involvement of the large scale circuits containing the devices that induce the memristive behavior, such as resistive switching memories. However, all existing memristive devices have a limitations in terms of the number of possible memristive states. 
For example, the analog implementations of the memristive pattern matcher and low voltage amplifier involving the memristors that replace the resistors in the circuits \cite{tcad} are practically very difficult to implement with the existing memristive devices. As the implementation of these circuits require the memristors to be programmed to exact level of resistance, the implementation of such circuits with the existing two state memristive devices is complex. One of the possibility is to use the set of memristive devices with $R_{on}$ and $R_{off}$ states connected in parallel and in series. However, the real behavior of such configuration should be tested considering {parasitics, leakage and sneak currents}, because the parallel and series connection of the memristors is different than in the resistors. The current flow in memristors should be considered. 
Memristive analog HTM implementation can involve the resistive switching memories that have been used for the implementations of the neural networks \cite{7727299}, however the HTM behavior will not be accurate due to the limited number of memristive states. 

{ One of the possibility to achieve multiple resistive states is to use $GST$ memristors with 16 resistive levels \cite{kuzum2011nanoelectronic}. However, the area of such memristors is larger and  $GST$ memristor technology is not always compatible with the CMOS technology. The possibility to achieved 64 resistive levels using $SiO_x$-based memristor with $Si$ diodes is illustrated in \cite{chang2016demonstration}. The overall size of such device is $41\times 19 \mu m^2 + 21.9 \times 21.9 \mu m^2$. However, the set and reset voltage of such device is very large for the HTM applications and designed circuits, because the minimum reset voltage can reach up to $17V$ with the time pulse of $1000 \mu s$.
The recent works illustrate the possibility of implementation of multilevel states with the use $SiN_x$-based memristor, which is compatible with the CMOS technology, can be used in a crossbar array and suppresses the issue of sneak path currents \cite{kim2017analog}. The switching time of such memristors is $100ns-200ns$, which is relatively fast comparing to HP $TiO_2$ memristors \cite{strukov2008missing}. However, the switching voltage is large (greater than $5V$ \cite{kim2017analog}). In addition, the memristor has not been tested for HTM applications. The HTM circuits have to be adjusted to use with $SiN_x$-based memristors, because the read pulse for HTM circuits was set to $1V$, while the read pulse for $SiN_x$-based memristors is $0.2V$ \cite{kim2017analog}. }

In addition, the real memristive devices are not always accurate and involve the problem of switching stochasticity. Two real memristors of the same technology can react differently to the same voltage applied across the memristor for the switching \cite{al2015memristors}. The switching time and level may vary or,  according to the theories of probabilistic behavior of the memristor, the switching may not occur. Such behavior of memristive devices may effect the performance accuracy of the HTM SP and HTM TM significantly. For example, in the HTM TM more training images and training cycles may require to achieve the same accuracy and system level HTM performance due to these variabilities in the behavior of the memristor.


\section{Conclusion}
\label{s6}

In this work, the comprehensive survey on the memristive HTM has been presented. The existing HTM digital, analog and mixed signal implementations have been reviewed and their advantages and drawbacks have been identified. The architectures of memristive HTM have been compared in terms of accuracy, on-chip area and power. In addition, the applications of HTM, including face, speech and handwritten digits recognition, gender classification and NLP, and proposed hardware solutions were reviewed. 

The analog memristive HTM implementation has an advantage in terms of processing speed, small on-chip area and possibility to adjust to total power consumption of the HTM circuit by replacing the conventional elements with low power circuit configurations. However, the analog  hardware limitations do not allow the implementation of the traditional HTM algorithm and impose several modifications of the original HTM algorithm. The open problems in the memristive HTM implementation include the scalability of the analog circuits and memristive crossbars, the limitation in the number of states and unpredictable behavior of real memristive devices and a lack of full {circuit level implementation} of the robust analog memory unit and learning algorithm for the weight update process in the HTM SP and TM.





\ifCLASSOPTIONcaptionsoff
  \newpage
\fi

\bibliographystyle{IEEEtran}
\bibliography{reference}

\begin{thebibliography}{10}
\providecommand{\url}[1]{#1}
\csname url@samestyle\endcsname
\providecommand{\newblock}{\relax}
\providecommand{\bibinfo}[2]{#2}
\providecommand{\BIBentrySTDinterwordspacing}{\spaceskip=0pt\relax}
\providecommand{\BIBentryALTinterwordstretchfactor}{4}
\providecommand{\BIBentryALTinterwordspacing}{\spaceskip=\fontdimen2\font plus
\BIBentryALTinterwordstretchfactor\fontdimen3\font minus
  \fontdimen4\font\relax}
\providecommand{\BIBforeignlanguage}[2]{{%
\expandafter\ifx\csname l@#1\endcsname\relax
\typeout{** WARNING: IEEEtran.bst: No hyphenation pattern has been}%
\typeout{** loaded for the language `#1'. Using the pattern for}%
\typeout{** the default language instead.}%
\else
\language=\csname l@#1\endcsname
\fi
#2}}
\providecommand{\BIBdecl}{\relax}
\BIBdecl

\bibitem{doremalen2008spoken}
J.~v. Doremalen and L.~Boves, ``Spoken digit recognition using a hierarchical
  temporal memory,'' in \emph{Ninth Annual Conference of the International
  Speech Communication Association}, 2008.

\bibitem{maltoni2011pattern}
D.~Maltoni, ``Pattern recognition by hierarchical temporal memory,'' 2011.

\bibitem{mattsson2011fruit}
O.~Mattsson, ``Fruit recognition by hierarchical temporal memory,'' 2011.

\bibitem{csapo2007object}
A.~B. Csap{\'o}, P.~Baranyi, and D.~Tikk, ``Object categorization using
  vfa-generated nodemaps and hierarchical temporal memories,'' in
  \emph{Computational Cybernetics, 2007. ICCC 2007. IEEE International
  Conference on}.\hskip 1em plus 0.5em minus 0.4em\relax IEEE, 2007, pp.
  257--262.

\bibitem{perea2009application}
A.~J. Perea, J.~E. Mero{\~n}o, and M.~J. Aguilera, ``Application of
  numenta{\textregistered} hierarchical temporal memory for land-use
  classification,'' \emph{South African Journal of Science}, vol. 105, no.
  9-10, pp. 370--375, 2009.

\bibitem{hawkinsintelligence}
J.~Hawkins and S.~Blakeslee, ``On intelligence. 2004,'' \emph{New York St.
  Martin’s Griffin}, pp. 156--8.

\bibitem{bami2016}
\BIBentryALTinterwordspacing
J.~Hawkins, S.~Ahmad, S.~Purdy, and A.~Lavin, ``Biological and machine
  intelligence (bami),'' 2016, initial online release 0.4. [Online]. Available:
  \url{http://numenta.com/biological-and-machine-intelligence/}
\BIBentrySTDinterwordspacing

\bibitem{george2005hierarchical}
D.~George and J.~Hawkins, ``A hierarchical bayesian model of invariant pattern
  recognition in the visual cortex,'' in \emph{Neural Networks, 2005. IJCNN'05.
  Proceedings. 2005 IEEE International Joint Conference on}, vol.~3.\hskip 1em
  plus 0.5em minus 0.4em\relax IEEE, 2005, pp. 1812--1817.

\bibitem{zyarah2015design}
A.~M. Zyarah, ``Design and analysis of a reconfigurable hierarchical temporal
  memory architecture,'' Master's thesis, Rochester Institute of Technology, 1
  Lomb Memorial Dr, Rochester, NY 14623, 6 2015.

\bibitem{ibrayevdesign}
T.~Ibrayev, A.~P. James, C.~Merkel, and D.~Kudithipudi, ``A design of htm
  spatial pooler for face recognition using memristor-cmos hybrid circuits,''
  in \emph{2016 IEEE International Symposium on Circuits and Systems (ISCAS)},
  May 2016, pp. 1254--1257.

\bibitem{fedorova2016htm}
A.~P. James, I.~Fedorova, T.~Ibrayev, and D.~Kudithipudi, ``Htm spatial pooler
  with memristor crossbar circuits for sparse biometric recognition,''
  \emph{IEEE Transactions on Biomedical Circuits and Systems}, vol.~PP, no.~99,
  pp. 1--12, 2017.

\bibitem{tcad}
O.~Krestinskaya, T.~Ibrayev, and A.~P. James, ``Hierarchical temporal memory
  features with memristor logic circuits for pattern recognition,'' \emph{IEEE
  Transactions on Computer-Aided Design of Integrated Circuits and Systems},
  vol.~PP, no.~99, pp. 1--1, 2017.

\bibitem{james2017design}
A.~James, A.~Irmanova, and T.~Ibrayev, ``Design of discrete-level memristive
  circuits for hierarchical temporal memory based spatio-temporal data
  classification system,'' \emph{IET Cyber-Physical Systems: Theory \&
  Applications}, 2017.

\bibitem{fan2016hierarchical}
D.~Fan, M.~Sharad, A.~Sengupta, and K.~Roy, ``Hierarchical temporal memory
  based on spin-neurons and resistive memory for energy-efficient
  brain-inspired computing,'' \emph{IEEE transactions on neural networks and
  learning systems}, vol.~27, no.~9, pp. 1907--1919, 2016.

\bibitem{streat2016non}
L.~Streat, D.~Kudithipudi, and K.~Gomez, ``Non-volatile hierarchical temporal
  memory: Hardware for spatial pooling,'' \emph{arXiv preprint
  arXiv:1611.02792}, 2016.

\bibitem{poster}
A.~Irmanova and A.~P. James, ``Htm sequence memory for language processing,''
  in \emph{Poster session presented at IEEE International Conference on
  Rebooting Computing (ICRC 2017)}, Nov 2017.

\bibitem{25}
N.~Inc., ``Hierarchical temporal memory including htm cortical learning
  algorithms,'' Tech. Rep., 2006.

\bibitem{spnew}
Y.~Cui, S.~Ahmad, and J.~Hawkins, ``The htm spatial pooler: a neocortical
  algorithm for online sparse distributed coding,'' \emph{bioRxiv}, p. 085035,
  2017.

\bibitem{mnatzaganian2017mathematical}
J.~Mnatzaganian, E.~Fokou{\'e}, and D.~Kudithipudi, ``A mathematical
  formalization of hierarchical temporal memory’s spatial pooler,''
  \emph{Frontiers in Robotics and AI}, vol.~3, p.~81, 2017.

\bibitem{tmnew}
Y.~Cui, S.~Ahmad, and J.~Hawkins, ``Continuous online sequence learning with an
  unsupervised neural network model,'' \emph{Neural computation}, 2016.

\bibitem{manem2011read}
H.~Manem and G.~S. Rose, ``A read-monitored write circuit for 1t1m multi-level
  memristor memories,'' in \emph{Circuits and systems (ISCAS), 2011 IEEE
  international symposium on}.\hskip 1em plus 0.5em minus 0.4em\relax IEEE,
  2011, pp. 2938--2941.

\bibitem{jung2012two}
C.-M. Jung, J.-M. Choi, and K.-S. Min, ``Two-step write scheme for reducing
  sneak-path leakage in complementary memristor array,'' \emph{IEEE
  Transactions on Nanotechnology}, vol.~11, no.~3, pp. 611--618, 2012.

\bibitem{lazzaro1988winner}
J.~Lazzaro, S.~Ryckebusch, M.~A. Mahowald, and C.~A. Mead, ``Winner-take-all
  networks of o (n) complexity,'' CALIFORNIA INST OF TECH PASADENA DEPT OF
  COMPUTER SCIENCE, Tech. Rep., 1988.

\bibitem{yakopcic2013memristor}
C.~Yakopcic, T.~M. Taha, G.~Subramanyam, and R.~E. Pino, ``Memristor spice
  model and crossbar simulation based on devices with nanosecond switching
  time,'' in \emph{Neural Networks (IJCNN), The 2013 International Joint
  Conference on}.\hskip 1em plus 0.5em minus 0.4em\relax IEEE, 2013, pp. 1--7.

\bibitem{ibrayev2017chip}
T.~Ibrayev, U.~Myrzakhan, O.~Krestinskaya, A.~Irmanova, and A.~P. James,
  ``On-chip face recognition system design with memristive hierarchical
  temporal memory,'' \emph{arXiv preprint arXiv:1709.08184}, 2017.

\bibitem{5937942}
H.~Abdalla and M.~D. Pickett, ``Spice modeling of memristors,'' in \emph{2011
  IEEE International Symposium of Circuits and Systems (ISCAS)}, May 2011, pp.
  1832--1835.

\bibitem{biol}
D.~Biolek, Z.~Kolka, V.~Biolkova, and Z.~Biolek, ``Memristor models for spice
  simulation of extremely large memristive networks,'' in \emph{2016 IEEE
  International Symposium on Circuits and Systems (ISCAS)}, May 2016, pp.
  389--392.

\bibitem{ascoli2015art}
A.~Ascoli, R.~Tetzlaff, Z.~Biolek, Z.~Kolka, V.~Biolkov{\`a}, and D.~Biolek,
  ``The art of finding accurate memristor model solutions,'' \emph{IEEE Journal
  on Emerging and Selected Topics in Circuits and Systems}, vol.~5, no.~2, pp.
  133--142, 2015.

\bibitem{MRTL}
A.~K. Maan, D.~A. Jayadevi, and A.~P. James, ``A survey of memristive threshold
  logic circuits,'' \emph{IEEE Transactions on Neural Networks and Learning
  Systems}, vol.~28, no.~8, pp. 1734--1746, Aug 2017.

\bibitem{georgia}
A.~James, T.~Ibrayev, and O.~Krestinskaya, ``Design and implication of a rule
  based weight sparsity module in htm spatial pooler,'' in \emph{Electronics ,
  Circuits and Systems (ICECS), 2017 24th IEEE International}.\hskip 1em plus
  0.5em minus 0.4em\relax IEEE, 2017.

\bibitem{krestinskaya2018feature}
O.~Krestinskaya and A.~P. James, ``Feature extraction without learning in an
  analog spatial pooler memristive-cmos circuit design of hierarchical temporal
  memory,'' \emph{Analog Integrated Circuits and Signal Processing}, pp. 1--9,
  2018.

\bibitem{vourkas2016emerging}
I.~Vourkas and G.~C. Sirakoulis, ``Emerging memristor-based logic circuit
  design approaches: A review,'' \emph{IEEE Circuits and Systems Magazine},
  vol.~16, no.~3, pp. 15--30, 2016.

\bibitem{martinez1998ar}
A.~Mart{\i}nez and R.~Benavente, ``The ar face database,'' \emph{Rapport
  technique}, vol.~24, 1998.

\bibitem{47}
R.~Senthilkumar and R.~K. Gnanamurthy, ``A detailed survey on 2d and 3d still
  face and face video databases part i,'' in \emph{Communications and Signal
  Processing (ICCSP), 2014 International Conference on}, April 2014, pp.
  1405--1409.

\bibitem{46}
R.~Ahdid, S.~Safi, and B.~Manaut, ``Approach of facial surfaces by contour,''
  in \emph{Multimedia Computing and Systems (ICMCS), 2014 International
  Conference on}, April 2014, pp. 465--468.

\bibitem{garofolo1993darpa}
J.~S. Garofolo, L.~F. Lamel, W.~M. Fisher, J.~G. Fiscus, and D.~S. Pallett,
  ``Darpa timit acoustic-phonetic continous speech corpus cd-rom. nist speech
  disc 1-1.1,'' \emph{NASA STI/Recon Technical Report N}, vol.~93, 1993.

\bibitem{lenc2015unconstrained}
L.~Lenc and P.~Kr{\'a}l, ``Unconstrained facial images: Database for face
  recognition under real-world conditions,'' in \emph{Mexican International
  Conference on Artificial Intelligence}.\hskip 1em plus 0.5em minus
  0.4em\relax Springer, 2015, pp. 349--361.

\bibitem{abdalla2011spice}
H.~Abdalla and M.~D. Pickett, ``Spice modeling of memristors,'' in
  \emph{Circuits and Systems (ISCAS), 2011 IEEE International Symposium
  on}.\hskip 1em plus 0.5em minus 0.4em\relax IEEE, 2011, pp. 1832--1835.

\bibitem{melis}
W.~J.~C. Melis, S.~Chizuwa, and M.~Kameyama, ``Evaluation of hierarchical
  temporal memory for a real world application,'' in \emph{Innovative
  Computing, Information and Control (ICICIC), 2009 Fourth International
  Conference on}, Dec 2009, pp. 144--147.

\bibitem{padilla2013performance}
D.~E. Padilla, R.~Brinkworth, and M.~D. McDonnell, ``Performance of a
  hierarchical temporal memory network in noisy sequence learning,'' in
  \emph{Computational Intelligence and Cybernetics (CYBERNETICSCOM), 2013 IEEE
  International Conference on}.\hskip 1em plus 0.5em minus 0.4em\relax IEEE,
  2013, pp. 45--51.

\bibitem{zidan2014leakage}
M.~A. Zidan, A.~Sultan, H.~A. Fahmy, and K.~N. Salama, ``Leakage analysis of
  crossbar memristor arrays,'' in \emph{Cellular Nanoscale Networks and their
  Applications (CNNA), 2014 14th International Workshop on}.\hskip 1em plus
  0.5em minus 0.4em\relax IEEE, 2014, pp. 1--2.

\bibitem{7727299}
Y.~Long, E.~M. Jung, J.~Kung, and S.~Mukhopadhyay, ``Reram crossbar based
  recurrent neural network for human activity detection,'' in \emph{2016
  International Joint Conference on Neural Networks (IJCNN)}, July 2016, pp.
  939--946.

\bibitem{kuzum2011nanoelectronic}
D.~Kuzum, R.~G. Jeyasingh, B.~Lee, and H.-S.~P. Wong, ``Nanoelectronic
  programmable synapses based on phase change materials for brain-inspired
  computing,'' \emph{Nano letters}, vol.~12, no.~5, pp. 2179--2186, 2011.

\bibitem{chang2016demonstration}
Y.-F. Chang, B.~Fowler, Y.-C. Chen, F.~Zhou, C.-H. Pan, T.-C. Chang, and J.~C.
  Lee, ``Demonstration of synaptic behaviors and resistive switching
  characterizations by proton exchange reactions in silicon oxide,''
  \emph{Scientific reports}, vol.~6, p. 21268, 2016.

\bibitem{kim2017analog}
S.~Kim, H.~Kim, S.~Hwang, M.-H. Kim, Y.-F. Chang, and B.-G. Park, ``Analog
  synaptic behavior of a silicon nitride memristor,'' \emph{ACS applied
  materials \& interfaces}, vol.~9, no.~46, pp. 40\,420--40\,427, 2017.

\bibitem{strukov2008missing}
D.~B. Strukov, G.~S. Snider, D.~R. Stewart, and R.~S. Williams, ``The missing
  memristor found,'' \emph{nature}, vol. 453, no. 7191, p.~80, 2008.

\bibitem{al2015memristors}
M.~Al-Shedivat, R.~Naous, G.~Cauwenberghs, and K.~N. Salama, ``Memristors
  empower spiking neurons with stochasticity,'' \emph{IEEE Journal on Emerging
  and Selected Topics in Circuits and Systems}, vol.~5, no.~2, pp. 242--253,
  2015.

\end{thebibliography}

\end{document}